\documentclass[usenatbib]{mnras}

\usepackage{amssymb}	

\usepackage{newtxtext,newtxmath}

\usepackage[T1]{fontenc}
\usepackage[normalem]{ulem}

\DeclareRobustCommand{\VAN}[3]{#2}
\let\VANthebibliography\thebibliography
\def\thebibliography{\DeclareRobustCommand{\VAN}[3]{##3}\VANthebibliography}

\usepackage{graphicx}	
\usepackage{amsmath}	

\usepackage{fontawesome5}
\usepackage{color}
\usepackage{xcolor}

\usepackage{lineno}

\newcommand{\Rfivehc}{R_{\rm 500c}}
\newcommand{\Mfivehc}{M_{\rm 500c}}
\newcommand{\Rtwohc}{R_{\rm 200c}}
\newcommand{\Mtwohc}{M_{\rm 200c}}
\newcommand{\Rtwohm}{R_{\rm 200m}}
\newcommand{\Mtwohm}{M_{\rm 200m}}

\newcommand{\MeanY}{\langle y \rangle}

\newcommand{\Rshock}{R_{\rm sh}}

\newcommand{\Rsp}{R_{\rm sp}}

\newcommand{\eg}{{\sl e.g.}, }        

\newcommand{\msol}{\ensuremath{\, {\rm M}_\odot}}    
\newcommand{\msun}{\ensuremath{\, {\rm M}_\odot}} 
\newcommand{\kpc}{\ensuremath{\, {\rm kpc}}}         
\newcommand{\mpc}{\ensuremath{\, {\rm Mpc}}}

\newcommand{\dln}{\ensuremath{{\rm d \ln}}}

\newcommand{\Rad}{R/\Rtwohm}
\newcommand{\logder}{\frac{\dln y}{\dln R}}

\newcommand{\citepinprep}[1]{(\textcolor{blue}{#1 et. al, in prep.})}

\definecolor{orcidlogocol}{HTML}{A6CE39}
\definecolor{purple}{RGB}{128, 0, 128}

\newcommand{\OrcidID}[1]{ \href[urlcolor = red]{https://orcid.org/#1}{\textcolor{lightgray}{\faOrcid}}}
\newcommand{\OrcidIDName}[2]{\href{https://orcid.org/#1}{#2}}

\newcommand*{\vcenteredhbox}[1]{\begingroup
\setbox0=\hbox{#1}\parbox{\wd0}{\box0}\endgroup}

\defcitealias{Anbajagane2022Shocks}{A22}

\title[Cosmological shocks around galaxy clusters]{Cosmological shocks around galaxy clusters: A coherent investigation with DES, SPT \& ACT}

\usepackage{eso-pic}
\AddToShipoutPictureBG*{%
  \AtPageUpperLeft{%
    \hspace{0.75\paperwidth}%
    \raisebox{-3.5\baselineskip}{%
      \makebox[0pt][l]{\textnormal{DES-2023-0784}}
}}}%

\AddToShipoutPictureBG*{%
  \AtPageUpperLeft{%
    \hspace{0.75\paperwidth}%
    \raisebox{-4.5\baselineskip}{%
      \makebox[0pt][l]{\textnormal{FERMILAB-PUB-23-525-PPD
}}
}}}%

\author[DES, SPT, \& ACT Collaboration]{
\parbox{\textwidth}{
\Large
\OrcidIDName{0000-0003-3312-909X}{D. Anbajagane}\thanks{Corresponding author email: dhayaa@uchicago.edu}(\vcenteredhbox{\includegraphics[height=1.2\fontcharht\font`\B]{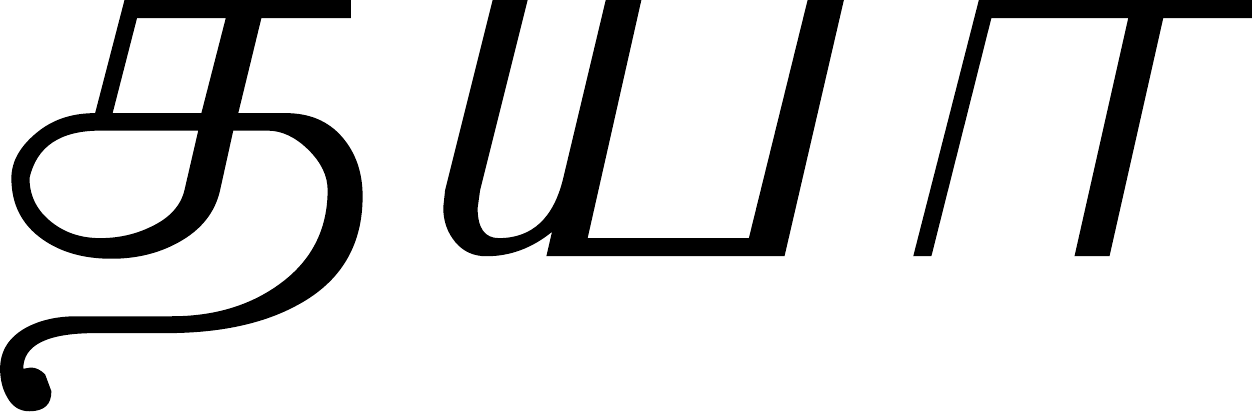}}),$^{1,\,2}$
\OrcidIDName{0000-0002-7887-0896}{C. Chang},$^{1,\,2}$
E.~J.~Baxter,$^{3}$
S.~Charney,$^{4}$
M.~Lokken,$^{5,6,7}$
M.~Aguena,$^{8}$
S.~Allam,$^{9}$
O.~Alves,$^{10}$
A.~Amon,$^{11,12}$
R.~An,$^{13}$
F.~Andrade-Oliveira,$^{10}$
% ...
}
}

\date{Accepted XXX. Received YYY; in original form ZZZ}

\pubyear{2021}

\begin{document}
\label{firstpage}
\pagerange{\pageref{firstpage}--\pageref{lastpage}}
\maketitle

\begin{abstract}
    We search for signatures of cosmological shocks in gas pressure profiles of galaxy clusters using the cluster catalogs from three surveys: the Dark Energy Survey (DES) Year 3, the South Pole Telescope (SPT) SZ survey, and the Atacama Cosmology Telescope (ACT) data releases 4, 5, and 6, and using thermal Sunyaev-Zeldovich (SZ) maps from SPT and ACT. The combined cluster sample contains around $10^5$ clusters with mass and redshift ranges $10^{13.7} < \Mtwohm/\msol < 10^{15.5}$ and $0.1 < z < 2$, and the total sky coverage of the maps is $
    \approx 15,000 \deg^2$. We find a clear pressure deficit at $R/\Rtwohm \approx 1.1$ in SZ profiles around both ACT and SPT clusters, estimated at $6\sigma$ significance, which is qualitatively consistent with a shock-induced thermal non-equilibrium between electrons and ions. The feature is not as clearly determined in profiles around DES clusters. We verify that measurements using SPT or ACT maps are consistent across all scales, including in the deficit feature. The SZ profiles of optically selected and SZ-selected clusters are also consistent for higher mass clusters. Those of less massive, optically selected clusters are suppressed on small scales by factors of 2-5 compared to predictions, and we discuss possible interpretations of this behavior. An oriented stacking of clusters --- where the orientation is inferred from the SZ image, the brightest cluster galaxy, or the surrounding large-scale structure measured using galaxy catalogs --- shows the normalization of the one-halo and two-halo terms vary with orientation. Finally, the location of the pressure deficit feature is statistically consistent with existing estimates of the splashback radius.
\end{abstract}


\section{Introduction}

Cosmological shocks are violent, high-energy phenomena that are a natural consequence of cosmic structure formation, and form in the far outskirts of massive, collapsed objects like galaxy clusters. They impact astrophysical processes like cosmic ray production and galaxy evolution, and are generated when colder gas is accreted onto a halo. The gravitational infall velocity of the cold gas will generically exceed the sound speed of the gas, especially for infall around massive halos, and this results in a high Mach number shock \citep[$M \sim 100$, \eg][]{Molnar2009ShocksInSZ}

The presence of such shocks impacts a wide array of astrophysical processes. These shocks are a natural thermodynamic boundary around the cluster, at the interface between the cluster-dominated gas component and the surrounding large-scale structure. They thereby also set the boundary within which the cluster has a thermodynamic impact on objects, such as galaxy quenching via ram-pressure stripping \citep[\eg][]{Zinger2016QuenchingShocks, Boselli2021RPSReview}. Shocks are sites for accelerating cosmic ray electrons via Diffusive Shock Acceleration \citep{Drury1983DSA, Blandford1987ShocksCRs}, and such accelerated cosmic ray electrons form a non-thermal tail in the energy distribution of the electron population \citep{Miniati2001NonThermal, Ryu2003CosmologicalShockWaves, Brunetti2014ClusterCRsReview}. The radial location of shock features also depends on the mass accretion rate of the cluster and can potentially serve as an observational proxy for the same \citep{Lau2015GasProfileOutskirts, Shi2016ShockMAR, Zhang2020MergerAcceleratedShocks, Zhang2021SplashShock}. The mass accretion rate has strong theoretical connections to key dark matter halo properties such as the concentration and formation time \citep{Wechsler2002Concentrations}, and has significant correlations with a wider range of halo properties \citep[\eg][]{Lau2021DMCorrelations, Anbajagane2022BaryImprint, Shin2023Splashcorr}. However, it has remained difficult to infer observationally.

This process of shock heating generates a thermal non-equilibrium between the electrons and ions, which can alter the expected thermodynamic profiles and will consequently need to be considered in analyses that include these cluster outskirts \citep{Fox1997ElectronNE, Ettori1998ElectronNE, Wong2009ElectronNE, Rudd2009ElectronNE, Akahori2010ElectronNE, Avestruz2015ElectronNE, Vink2015ElectronNE}. Specifically, shocks preferentially heat ions over electrons given the mass difference of the two species, and at the low-number densities of the cluster outskirts, the two species may not interact often enough to equilibrate. This will lead to a deficit in the measured SZ profiles --- which traces the \textit{electron}, not ion, temperature --- near a shock, and such a deficit has been observed previously with SPT data \citep{Anbajagane2022Shocks}. 
In addition, an accurate model of these cluster outskirts --- particularly near the transition regime between the bound component and the large-scale structure --- will be beneficial for studies of the large-scale gas pressure fields \citep[\eg][]{Hill2013tSZPowerSpec, Horowitz2017tSZCosmology, Tanimura2021PlancktSZCosmo} as well as cross-correlations of the gas pressure with galaxy and galaxy cluster positions \citep[\eg][]{Hajian2013PlanckWMAPxROSAT, Vikram2017GalaxyGroupstSZ, Hill2018tSZxGroups, Pandey2019GalaxytSZ, Pandey2020tSZCrossForecast, Sanchez2023DESxSPT},
with weak-lensing shears \citep[\eg][]{Ma2015PlanckxCFHTLens, Hojjati2017PlanckxRCSLenS, Osato2018PlanckxRCSLenS, Osato2020PlanckxHSC, Shirasaki2020XraySZLensing, Gatti2021DESxACT, Pandey2021DESxACT}, or with X-ray luminosity \citep{Shirasaki2020XraySZLensing}; these kinds of studies are positioned to provide strong and complementary constraints on astrophysical, as well as cosmological, processes. The model will also be beneficial for understanding the impact from the gas dynamics of the outskirts on the weak lensing signal (via the impact of gas dynamics on the total matter field) --- this impact is a significant limitation in extracting cosmological information from the lensing signal \citep[\eg][]{Gatti2020Moments, Krause2021Methods, Secco2022Shear, Amon2022Shear, Anbajagane2023Inflation, Anbajagane2023CDFs} --- and for subsequently modelling the impact via a halo-model approach \citep[\eg][]{Schneider2019Baryonification, Chen2023BaryonificationDES}.

While a wide variety of physical processes are influenced by the presence of shocks, the cosmological shocks are themselves simple, as their formation has two basic requirements: a matter component that is collisional and thus behaves hydrodynamically (``gas''), and an influx of this collisional matter onto a halo via gravitational infall. However, both hydrodynamics and gravitational infall are highly asymmetric processes with complicated geometries, and so, in practice, these shocks have a rich phenomenology with intricate, subtle behaviors.

This phenomenology has been extensively studied in simulations over the past many decades. The first studies used non-radiative simulations with gas dynamics but no astrophysical processes \citep{Quilis1998ShocksSims, Miniati2000ShocksSims, Ryu2003CosmologicalShockWaves, Skillman2008ShocksSims, Molnar2009ShocksInSZ, Hong2014ShockSims, Hong2015ShocksSims, Schaal2015ShocksIllustrisNR}. These were then followed by studies using simulations that include gas cooling and star formation \citep{Vazza2009ShocksSims, Planelles2013ShocksSims, Lau2015GasProfileOutskirts, Nelson2016ShocksSimsMW, Aung2020SplashShock}, and also include the effects of feedback from supernovae and active galactic nuclei \citep{Kang2007ShocksSims, Vazza2013FeedbackShocksSims, Vazza2014ShocksSims, Schaal2016HydroShocksIllustris, Baxter2021ShocksSZ, Planelles2021ShocksSims, Baxter2023C200cSZ, Sayers2023PressureThe300}. Some works have also opted to model the evolution of cosmic-rays --- which are generated at the shocks --- alongside galaxy formation \citep{Pfrommer2007CRsImpactSZandXray}, while others employ idealized simulations to understand the propagation of shocks and their dependence on different merger events \citep{Pfrommer2006ShocksSims, Ha2018ShocksSims, Zhang2019MergerShock, Zhang2020MergerAcceleratedShocks, Zhang2021SplashShock}. A number of works have also theoretically estimated the potential signal-to-noise of shocks from various surveys/instruments \citep[\eg][]{Kocsis2005Shocks, Baxter2021ShocksSZ}.

In the current picture, cosmological shocks form at different radial locations around the galaxy cluster depending on the mechanism that generates them. The accretion of pristine cold gas --- which has a low sound speed and is primarily found in low-density regions such as cosmic voids --- onto the thermalized, bound gas component results in a shock of a high Mach number ($M \sim 100$) and discontinuities in the profiles of many thermodynamic quantities such as temperature, entropy, pressure, and density. This shock --- approximately located near the virial radius of the cluster --- is oftentimes referred to as an accretion shock \citep[\eg][]{Lau2015GasProfileOutskirts, Aung2020SplashShock, Baxter2021ShocksSZ} or an external shock \citep{Ryu2003CosmologicalShockWaves}, and has a theoretical foundation that goes back many decades \citep{Bertschinger1985SelfSimilar}. Closer to the cluster core, the supersonic infall of galaxies and gas clumps into the hot, ionized gas leads to a series of bow shocks with weak Mach numbers, that are referred to as internal shocks \citep{Ryu2003CosmologicalShockWaves}. Furthermore, \citet{Zhang2019MergerShock, Zhang2020MergerAcceleratedShocks} found that these bow shocks detach from the infalling substructure, leading to a runaway merger shock that then collides with the accretion shock. This generates a new shock, named the \textit{Merger-accelerated Accretion Shock} or MA-shock, that is both further out and longer lived than the original accretion shock. The infall of substructure is a common process during structure formation, and so most shocks observed in the cluster outskirts are expected to be MA-shocks and can have radial locations between $1 \lesssim R/\Rtwohm \lesssim 2.5$ depending on the accretion history of the cluster \citep{Zhang2021SplashShock}. These structures, given their origin in the large-scale accretion of matter, are connected to other features in the cluster outskirts such as the splashback radius \citep{Adhikari2014Splashback,Diemer2014Splashback}. This feature has been found in various datasets \citep{Baxter2017SplashbackSDSS, Chang2018SplashbackDES, Shin2019SplashbackDESxACTxSPT, Adhikari2020SplashbackCosmicClock, Shin2021SplashbackDESxACT} and its connection to cosmological shocks has been explored via both analytic calculations and simulations \citep{Shi2016ShockMAR, Aung2020SplashShock, Baxter2021ShocksSZ, Zhang2021SplashShock}.

While many simulation-based studies exist on the formation and evolution of these shocks, there are only a few observational studies of these features. A key observable for studying these shocks is the cluster gas pressure profiles, measured via the thermal Sunyaev-Zel'dovich (SZ) signature of clusters \citep{Sunyaev1972SZEffect}. The SZ effect is the inverse Compton scattering of cosmic microwave background (CMB) photons off energetic electrons in the hot intra-cluster medium \citep[see][for reviews]{Carlstrom2002SZReview, Mroczkowski2019SZreview}. While cluster thermodynamic properties have traditionally been studied using X-ray observations, the SZ effect has emerged as the more ideal probe for the cluster outskirts as its signal amplitude depends linearly with density, whereas for X-rays this dependence is quadratic. Many of the existing observational works --- using either X-ray or SZ --- do not explicitly focus on shocks and most are limited to small, often single, cluster samples at lower redshifts \citep{Akamatsu2011ElectronNEAbell, Akahori2012BulletClusterNE, Akamatsu2016ElectronNEXray, Basu2016ALMASZ, DiMascolo2019Shocks, DiMascolo2019ShocksBullet, Hurier2019ShocksSZPlanck, Pratt2021ShocksPlanck, Zhu2021ShockXray}. More general studies of gas thermodynamic profiles, without a specific focus on shocks, do not push beyond $r \gtrsim \Rfivehc$ \citep[\eg][]{McDonald2014SPTGasProfiles, Ghirardini2017ChandraPressureProfile, Romero2017MUSTANGPressure, Romero2018MultineProfile, Ghirardini2018ShockAbellXMM}, though some do exist \citep{Planck2013PressureProfiles, Sayers2013SZBolocam, Sayers2016BolocamPlanck, Amodeo2021ACTxBOSS, Schaan2021ACTxBOSS, Melin2023SZProfiles, Lyskova2023Xray3R500c}.

\citet{Anbajagane2022Shocks}, henceforth \citetalias{Anbajagane2022Shocks}, performed the first analysis of the cluster outskirts with a large statistical sample of $10^2 - 10^3$ clusters, and found evidence of a pressure deficit at the cluster virial radius. This work is a follow-up on \citetalias{Anbajagane2022Shocks}, and our goals are to: (i) to strengthen the evidence for the pressure deficit with additional, sensitive SZ data, (ii) compare the SZ profiles and their pressure deficit feature, between SZ-selected and optically selected cluster catalogs, and between measurements from different SZ maps (ACT and SPT), (iii) measure cluster profile outskirts for lower-mass clusters, $\Mtwohm < 10^{14.5} \msol$, (iv) extract anisotropic features of the profile outskirts, using SZ image shapes, the Brightest Cluster Galaxy (BCG) shapes, or the large-scale density field, and finally (v) compare the location of detected features with other physical cluster radii, namely the splashback radius. We achieve all of the above by expanding our study to include additional surveys: an optically selected cluster catalog from the Dark Energy Survey (DES) Year 3 dataset, and an SZ map from ACT Data Release (DR) 6. Both datasets were not used in the work of \citetalias{Anbajagane2022Shocks}. The availability of the ACT DR6 map also allows us to now use the full ACT DR5 cluster catalog, whereas \citetalias{Anbajagane2022Shocks} were limited to using a subset ($\approx 25\%$) of the catalog that overlapped with the ACT DR4 map.

We organize this work as follows: in \S \ref{sec:Data} we describe the survey datasets used in this work and in \S \ref{sec:Measurement_Modeling} our choices for the profile measurement procedure and the theoretical modelling. Our results on shocks are shown in \S \ref{sec:Results} and their connections to other large-scale structure features are explored in \S \ref{sec:StructureFormation}. We conclude in \S \ref{sec:Discussion_Conclusions}.

\section{Data} \label{sec:Data}

We use data from three wide-field surveys --- the Dark Energy Survey (DES) Year 3, the South Pole Telescope (SPT) SZ survey, and the Atacama Cosmology Telescope (ACT) Data releases 4, 5, and 6 --- to constrain the cluster pressure profile on large scales. In contrast to \citetalias{Anbajagane2022Shocks}, we do not consider profiles from the \textit{Planck} SZ map, though \textit{Planck} data are used in the construction of the ACT and SPT maps (described below in Section \ref{sec:SPT_Data} and Section \ref{sec:ACT_data}). The former choice is because the 10$\arcmin$ resolution of the \textit{Planck} SZ map (which is an order-of-magnitude larger than the 1$\arcmin$ resolution of SPT and ACT) is a limiting factor in detecting shock features. The \textit{Planck} cluster catalog \citep{Planck2016ClusterCatalog} also has significant overlap with the SPT and ACT catalogs used in this work; 45\% of the 1093 \textit{Planck} clusters are found within either the ACT or SPT footprints.

The clusters in our samples are labeled by their spherical overdensity mass, $\Mtwohm$, which is defined as,
\begin{equation} \label{eqn:SO_masses}
    M_{\rm \Delta} = \rho_\Delta\frac{4\pi}{3} R_{\rm \Delta}^3,
\end{equation}
with $\rho_\Delta = 200\rho_m(z)$, where $\rho_m(z)$ is the mean matter density of the Universe at a given epoch. The associated radius is denoted as $\Rtwohm$. Features at the cluster outskirts, such as shocks, follow a more self-similar evolution when normalized by this radius definition \citep{Diemer2014Splashback, Lau2015GasProfileOutskirts}.

Both SPT and ACT infer $\Mfivehc$ from the integrated tSZ emission around each cluster, while DES infers $\Mfivehc$ from the cluster richness, where richness is the probabilistic number of satellite galaxies in the cluster. We then convert the $\Mfivehc$ estimate into $\Mtwohc$ and $\Mtwohm$ using the concentration-mass relation from \citet{Diemer2019Concentration} and the publicly available routine from the \textsc{COLOSSUS}\footnote{\url{https://bdiemer.bitbucket.io/colossus/}} open-source python package \citep{Diemer2018COLOSSUS}. We find our results are insensitive to assuming other choices for the concentration-mass relation \citep[\eg][]{Child2018ConcentratioMassRelation, Ishiyama2020UchuuConcentration}. The impact of baryons on this relation is also negligible at these halo masses and so is not considered here \citep*[\eg][]{Beltz-Mohrmann2021BaryonImpactTNG, Anbajagane2022BaryImprint, Shao2022BaryonImprints, Shao2023BaryonImprints}. 
Both $\Mfivehc$ and $\Mtwohc$ are defined by Equation \ref{eqn:SO_masses} but with alternative density contrasts of $\rho_\Delta = 500\rho_c(z)$ and $\rho_\Delta = 200\rho_c(z)$, respectively. Here, $\rho_c(z)$ is the critical density of the Universe at a given epoch. The mass and redshift distributions of the different cluster samples are shown in Figure \ref{fig:Characterize_Dataset}.

The tSZ amplitude is reported as the dimensionless $y$ parameter,
\begin{equation}\label{eqn:tSZ_y_def}
    y \equiv \frac{k_B\sigma_T}{m_e c^2}\int n_e T_e dl,
\end{equation}
where $k_B$ is the Boltzmann constant, $\sigma_T$ is the Thomson cross-section, $m_e c^2$ is the rest energy of an electron, $n_e$ and $T_e$ are the electron number density and temperature, respectively, and $l$ is the physical line-of-sight distance. Thus $y$ represents the electron pressure integrated along the line-of-sight.

The tSZ effect corresponds to CMB photons scattering off electrons with a thermal (i.e. Maxwellian) energy/momentum distribution. There exist similar effects, called the relativistic SZ (rSZ) and non-thermal SZ (ntSZ), which correspond to photons scatteringoff electrons with non-Maxwellian energy distributions, and may leak into the measured tSZ signal \citep{Mroczkowski2019SZreview}. In the rSZ effect, the presence of high-temperature electrons ($T_{\rm e} \gtrsim 5\, {\rm keV}$) requires relativistic corrections to the procedure for making the SZ maps. These corrections, however, are $\lesssim 5\%$ \citep[see Figure 1]{Erler2018rSZ} and are subdominant to the amplitudes of the features discussed in our work.\footnote{The work of \citet{Lee2022rSZ} shows the rSZ effect in simulations scales self-similarly as $\propto M^{2/3}$, or alternatively $\propto Y^{2/5}$, and so given our cluster sample spans across an order-of-magnitude in mass, the rSZ effect would change at most by a factor of two across our cluster sample. Note, however, that this is a factor two difference in an effect that contributes $<5\%$ to the total signal.} The ntSZ effect can be generated by a cosmic ray electron population, but is a subdominant effect within $\Rtwohc$ of the cluster, where cosmic rays make up $\lesssim 1\%$ of the total pressure \citep{Ackermann2014CRpressure}. Beyond this radius, the cosmic ray energy fraction is not well constrained. For this work, we follow \citetalias{Anbajagane2022Shocks} in assuming the ntSZ continues to be subdominant in the outskirts, and point out that the features we discuss are unaffected even if the ntSZ contaminates the tSZ at the $10\%$ level.

\subsection{The Dark Energy Survey (DES) Year 3}
\label{sec:DES_Data}

DES Y3 is a 5000 $\deg^2$ photometric survey of the southern sky in five bands ($grizY$). Galaxy clusters are identified using the \textsc{RedMaPPer} algorithm \citep{Rykoff2014Redmapper}, which identifies clusters from overdensities of red-sequence galaxies. Each cluster is assigned a ``richness'', $\lambda$, which is analogous to the number of red galaxies in the cluster. \textsc{redMaPPer} assigns each galaxy $i$ a probability that it is a satellite of galaxy cluster $j$. The richness of cluster $j$ is then the sum of these probabilities. 

This richness is used alongside a richness--mass relation --- which can be calibrated using various methods such as galaxy lensing \citep{McClintock2018WLClusterMass}, CMB lensing \citep{Baxter2018CMBLensingClusterMass}, cross-correlations of probes \citep[]{To2021_6x2+NC}, galaxy velocity dispersion \citep{Farahi2016StackedSpectro, Anbajagane2022GalVelBias}, etc. --- to obtain a mass estimate for each cluster. In this work, we use the richness--mass relation from \citet[][see their Equation 16]{Costanzi2021DESxSPT}, which is calibrated using a combination of optical and SZ cluster measurements --- namely, the DES cluster number counts and the SPT observable-mass relation --- for clusters with $\lambda \geq 20$. The observable-mass relation was in turn calibrated with targeted weak-lensing measurements. Note that the catalogs we use have objects of lower richness ($\lambda \approx 10$) and thus the inferred mass of these objects could be biased given we must extrapolate the scaling relation of \citet{Costanzi2021DESxSPT} to this regime. There are no well-calibrated richness--mass relations in this regime, and thus extrapolation is a necessity. In Section \ref{sec:GroupScales} we discuss the impact of such mass biases in our analysis. 

We also use a cluster signal-to-noise ratio as a weight when averaging the profiles across the sample (see Section \ref{sec:Measurement}). For DES, this signal-to-noise is taken to be the ratio of the richness over the richness uncertainty, $\lambda/\Delta\lambda$, where richness and the uncertainty are taken from the \textsc{RedMaPPer} columns \texttt{LAMBDA\_CHISQ} and \texttt{LAMBDA\_CHISQ\_E}, respectively.

Finally, we also use two different galaxy samples to enable oriented stacking of the cluster profiles. First, we use the DES Y3 source galaxy shape catalog \citep{Gatti2021ShearCatalog} --- where the shapes were measured using the \textsc{Metacalibration} code \citep{Sheldon2017Metacal} --- to obtain the orientation of the BCG of each cluster.\footnote{We have verified that using alternative shape measurements, such as those from the single object fitting procedure \citep{Sevilla2021Y3Gold}, results in similar orientations for the galaxies.} Then, we also use the magnitude-limited lens galaxy catalog, \textsc{Maglim} \citep{Porredon2021Maglim}, to infer the density field in the DES footprint, from which we can estimate a cluster orientation based on large-scale structure. This follows the methods of \citet{Lokken2022Superclustering}, and is discussed further in Section \ref{sec:Filaments}. Both datasets are part of the publicly available DES Y3 data release.\footnote{\url{https://des.ncsa.illinois.edu/releases/y3a2}}

\subsection{The South Pole Telescope (SPT) SZ Survey} \label{sec:SPT_Data}

SPT-SZ is a $2500 \rm \, deg^2$ survey of the southern sky at 95, 150, and 220 GHz, and was conducted using the South Pole Telescope \citep{Carlstrom2011}. The SZ map used in our analysis was presented in \citet{Bleem2021SPTymap}, has an angular resolution of $1.25^\prime$, and is made using data from both SPT-SZ and the \textit{Planck} 2015 data release; the former provides lower-noise measurements of the small scales, whereas the latter does the same for larger scales (multipoles $\ell \lesssim 1000$). The \textit{Planck} data consists of the 100, 143, 217, and 353 GHz maps from the High Frequency Instrument (HFI). The SZ map is constructed with the Linear Combination (LC) algorithm \citep[see][for a review]{Delabrouille2009ReviewLCAlgorithm}, applied to the maps of different frequencies. The weights of the linear combination are chosen so as to minimize the total variance in the output map. The weights are also modified to reduce contamination from the cosmic infrared background (CIB); see Section 3.5 in \citet{Bleem2021SPTymap} for more details. In our analysis, the map is further masked to remove point sources as well as the top 5\% of map regions most dominated by galactic dust. This is done using the binary masks provided in \citet[][see point 4 in their Appendix A]{Bleem2021SPTymap}.

The galaxy cluster catalog from this data contains 516 clusters that were first identified in \citet{Bleem2015ClusterCatalogSPT}, and were assigned updated redshifts and mass estimates in \citet{Bocquet2019ClusterCatalogSPT}. We use the latter, updated catalog for our work, where the mass is estimated via a joint modeling of SZ, X-ray, and weak lensing measurements. Both the map and the cluster catalog are publicly available.\footnote{\url{https://lambda.gsfc.nasa.gov/product/spt/spt_prod_table.cfm}} Our masses come from the \texttt{M500} column and signal-to-noise ratio (SNR) from the \texttt{XI} column. This dataset is the exact same as the SPT-SZ data used in \citetalias{Anbajagane2022Shocks}.

\begin{figure}
    \centering
    \includegraphics[width = \columnwidth]{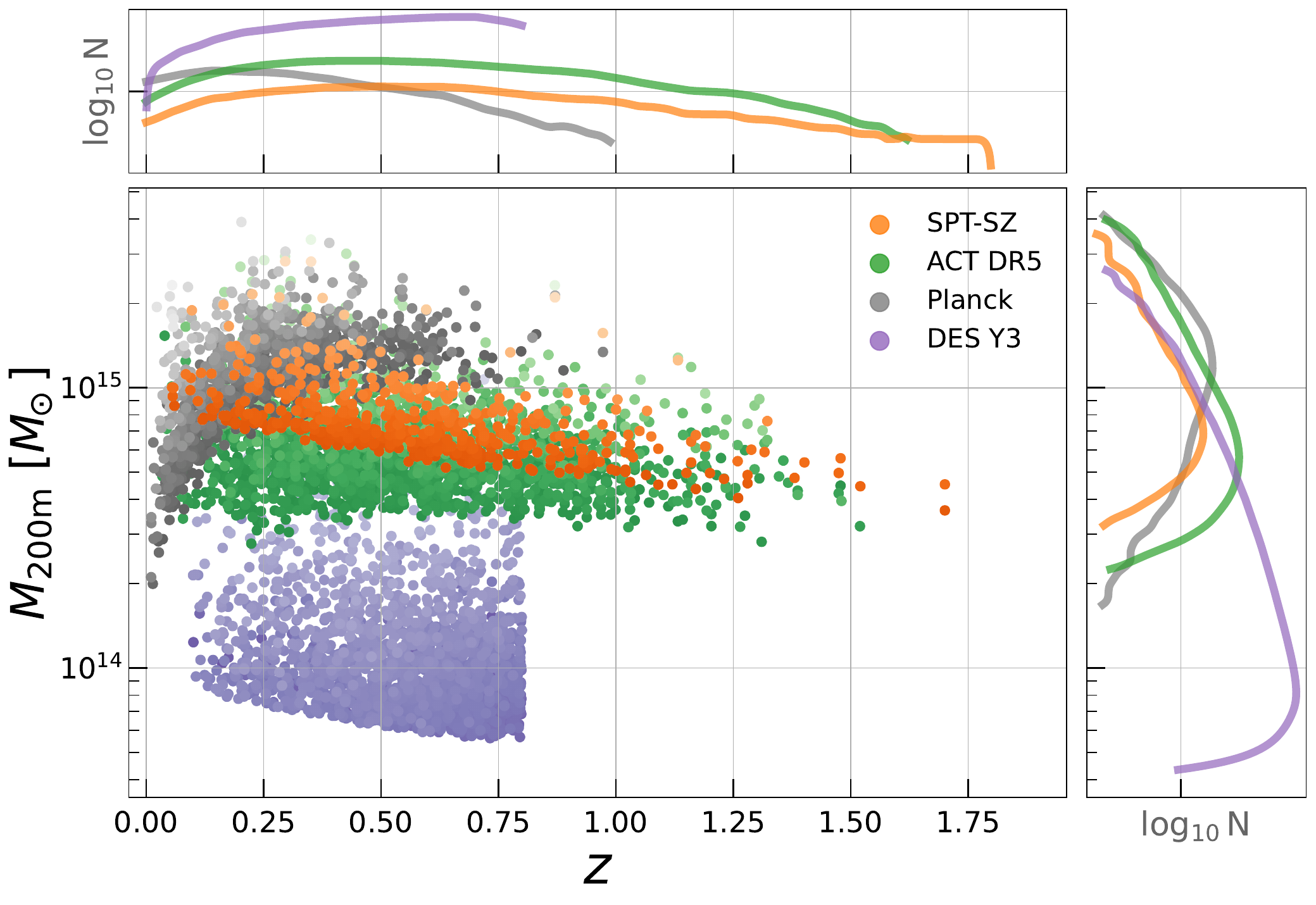}
    \caption{The mass-redshift plane of the cluster samples from SPT, ACT, and DES used in this work. The \textit{Planck} catalog is shown in grey for reference. The top and right panels show the 1D distributions for redshift and cluster mass, respectively. For visibility, we only plot a randomly chosen subset of DES clusters, with $N = 5000$. The 1D distributions are estimated using the full samples. The SPT and ACT samples have similar redshift distributions, with a median of $z \approx 0.55$, while DES Y3 is limited to $0.1 < z < 0.8$. DES also extends to much lower masses across all redshifts, where the masses are computed using the mass--richness relation of \citet[][see their Equation 16]{Costanzi2021DESxSPT}. The color tones of the points show $\log_{10}\rm SNR$, the signal-to-noise ratio of each cluster detection, with lighter colors indicating a higher SNR. The mean redshift and mass of the different samples are listed in Table \ref{tab:Results}.}
    \label{fig:Characterize_Dataset}
\end{figure}

\subsection{Atacama Cosmology Telescope (ACT) data releases 4, 5, and 6} \label{sec:ACT_data}

The ACT data covers 90, 150, and 220 GHz frequencies, and the maps from data release (DR) 6 cover $\approx 13,\!000 \deg^2$ of the sky (after applying the relevant masks; see discussion below). The SZ map \citep{Coulton2023ACTyMaps} has a resolution of $1.6^\prime$, and makes use of data from both ACT and the \textit{Planck} NPIPE data release  \citep{Planck2020NpipeMap}; as was the case with SPT, the former data inform small-scales and the latter, the large-scales ($\ell \lesssim 1000$). Note that the \textit{Planck} data here consist of eight frequency channels from 30 to 545 GHz, whereas the map from \citet{Bleem2021SPTymap} used four of these channels. The map is made using a \textit{Needlet Internal} Linear Combination (NILC) algorithm.

In our analysis, the map is further masked to remove point sources and dusty regions. The ACT DR6 mask is an apodized, continuous mask, not a binary one, and we continue with our aggressive masking by only selecting pixels for which the mask value is 1, meaning the impact of point sources and dust is negligible in this pixel. Note that this map does \textit{not} use the \textsc{HEALPix} pixelation scheme implemented in \textsc{Healpy} and instead uses the Plate Carr\'ee scheme implemented in \textsc{Pixell}\footnote{\url{https://pixell.readthedocs.io/en/latest/}}, a package optimized to work with partial sky maps in the flat-sky approximation. We use the ACT DR6 map in its native scheme and do not convert it to a \textsc{HEALPix} format.

We also use the $\approx 4200$ clusters from ACT DR5\footnote{\url{https://lambda.gsfc.nasa.gov/product/act/actpol_dr5_szcluster_catalog_info.html}} catalog \citep{Hilton2021ACTClusters}, which covers the same area as the ACT DR6 map. Note that only the subset of the ACT DR5 catalog, that corresponded to the 2000 $\deg^2$ area of the ACT DR4 map was used in \citetalias{Anbajagane2022Shocks}. The redshift distribution of the ACT DR5 cluster sample is similar to that of the SPT-SZ sample. As was the case in \citetalias{Anbajagane2022Shocks}, the cluster masses come from the \texttt{M500cCal} column described in \citet[see their Table 1]{Hilton2021ACTClusters}, which contains a weak lensing mass calibration factor. While other lensing-based calibrations also exist for the ACT data \citep[\eg][]{Roberston2023ACTMasses}, we use the fiducial calibration included in the catalog of \citet{Hilton2021ACTClusters}. The SPT and ACT masses are similar \citep[\eg][see their Section 5.1]{Hilton2021ACTClusters}, with the agreement at a level adequate for astrophysical analyses.

\section{Measurement and Modeling} \label{sec:Measurement_Modeling}

We first describe our procedure for measuring the stacked SZ profile in \S \ref{sec:Measurement}, and then in \S \ref{sec:Model_detection_significance} the theoretical halo model we compare the measurements with, including how we quantify the significance of any features in the data. 

\subsection{Measurement Procedure} \label{sec:Measurement}

Our measurement procedure closely follows that described in \citetalias{Anbajagane2022Shocks}, with some notable changes. We reproduce the main aspects of the measurement here for completeness but also point readers to \citetalias{Anbajagane2022Shocks} for a more detailed discussion on some elements of the procedure. Overall, the measurement procedure can be broken into four steps: (i) stacked profiles, (ii) logarithmic derivatives, (iii) bin-to-bin covariance matrix, and; (iv) feature locations.

\textbf{Estimating stacked profiles:} For each cluster, we compute the $\MeanY$ profile in 50 logarithmically spaced radial bins in the range $r \in [0.1, 20]\Rtwohm$. We convert between angular and physical scales using the angular diameter distance estimated at the redshift of each cluster. The profile also has a mean background value subtracted from it. Previously this background was estimated by measuring the average profile around uniform random points across the whole map. This method was adequate for maps with mostly homogeneous survey properties, but can cause biases for maps with inhomogeneous survey properties, such as ACT DR6 where some regions of the sky are observed to significantly higher depth than other regions. We have thus updated our background subtraction procedure to capture this inhomogeneity. We take the region spanned by the cluster catalog, and split it into different ``tiles'' based on \textsc{Healpix} pixelization of $\rm NSIDE = 4$. We have verified that our results below are robust if we instead use $\rm NSIDE = 8$ or $\rm NSIDE = 16$. We continue using $\rm NSIDE = 4$ for our analysis given it is computationally cheaper. Once we tile the maps, we estimate the background separately in each tile by measuring profiles around all random points in the chosen tile. During background subtraction for a given cluster, we choose the background profile of the tile closest to that cluster.\footnote{Alternatively, one could also produce a catalog of random points that sample the sky in a manner consistent with the cluster catalog of a given survey, and this can be produced by using maps of multiple survey properties. We have pursued our inhomogenous background subtraction method as it can be performed without requiring this additional data product.} Previously, all clusters had a common background profile subtracted from them, whereas now the subtracted profile varies across the sky. 

In \citetalias{Anbajagane2022Shocks}, we did not consider the contamination in a cluster's measured profile due to interloper clusters in the foreground/background. Interlopers are distant in physical, 3D space but appear close in projected, 2D space. We have explicitly checked this effect --- by masking out all potential interlopers when measuring the profiles of a given cluster --- and found it does not impact the features we discuss in this work. In our test, an interloper is defined as any cluster whose line-of-sight distance from the target cluster is $R > 20\Rtwohm$. An object with a large line-of-sight separation from a given cluster is not part of the latter's local large-scale environment but can appear so in projected 2D space where the line-of-sight separation is not relevant. Thus, selecting clusters where the line-of-sight  separation is greater than $20\Rtwohm$ isolates such interlopers. The choice of $20\Rtwohm$ is because that is the largest radius we measure the profiles to. We convert the cluster redshift to physical distance assuming a fiducial Lambda Cold Dark Matter ($\Lambda$CDM) cosmology with $\Omega_{\rm m} = 0.3$ and $h = 0.7$, and use the distances to identify the interlopers. Photometric redshift uncertainties and cluster line-of-sight peculiar velocities will affect the accuracy of the distance estimate. Even so, this test is useful as an approximate check of the interlopers' impact. For our main analysis below, we do not perform any interloper masking as we have confirmed it is a negligible effect.

The profiles of the individual clusters are then stacked, with each profile being weighted by the corresponding cluster's signal-to-noise (\texttt{SNR}). Performing a standard average/stack with no weights does not change the result (see Appendix A in \citetalias{Anbajagane2022Shocks}). Note that for a given cluster, any radial bin that did not have any pixels in it --- most commonly the case in the cores of high redshift clusters due to the limited angular resolution --- is masked, and thus ignored, during the stacking. 
The uncertainty of the stacked profile is obtained through a leave-one-out jackknife resampling. The $i$-th jackknife sample of the stacked profile can be written as
\begin{align} \label{eqn:Jackknife_Mean}
    \MeanY_i(r) =\,\, & \frac{1}{W_i(r)}\sum_{j \neq i}^{N_{\rm cl}} \,\,y_j(r) w_j \delta_j(r), \\
    W_i(r) =\,\, & \sum_{j \neq i}^{N_{\rm cl}} \,\, w_j \delta_j(r), 
\end{align}
where $w_j$ is the \texttt{SNR} per cluster used in the weighted average, $\delta_j(r)$ is 1 if the datapoint for radius $r$ in cluster $j$ is unmasked and 0 otherwise, $N_{\rm cl}$ is the total number of clusters.
In this notation, $\MeanY_i$ is the mean profile of the sample with cluster $i$ removed, and $y_j$ is the individual profile measurement from cluster $j$. The variance on the mean profile is then given by,
\begin{align} \label{eqn:Jackknife_Var}
    \sigma^2(r) =\,\,& \frac{N(r) - 1}{N(r)}\sum_{j = 1}^{N_{\rm cl}}  \,\,\bigg(\MeanY_j(r) - \Bar{\MeanY}(r)\bigg)^2\delta_j(r),\\
    N(r) =\,\, & \sum_{j = 1}^{N_{\rm cl}} \,\, \delta_j(r),
\end{align}
where $\Bar{\MeanY}$ is the mean of the distribution of jackknife estimates computed in Equation \eqref{eqn:Jackknife_Mean}. Note that Equation \eqref{eqn:Jackknife_Var} has an additional factor of $N - 1$ compared to the traditional definition of the variance, as required when using a jackknife estimator for the variance.

\textbf{Estimating logarithmic derivatives:} Shocks are generally characterized by sharp changes in thermodynamic quantities, and have been identified in some previous works as the point of steepest descent in the pressure profiles \citep[\eg][]{Aung2020SplashShock, Baxter2021ShocksSZ}. This corresponds to measuring minima in the logarithmic derivative. Derivatives, however, are affected by noise and we alleviate this by smoothing the stacked profiles with a Gaussian of width $\sigma_{\ln r} = 0.16$, which is $1.5$ times the logarithmic bin width, $\Delta \ln r \approx 0.11$. All profiles are smoothed by this scale, and we present results only for the range $0.3 < R/\Rtwohm < 10$ which does not contain any edge effects due to the smoothing. \citetalias{Anbajagane2022Shocks} (see their Appendix A) have already shown that smoothing choices have negligible impact on the final results.

The log-derivative of the smoothed mean profile is computed using a five-point method,
\begin{equation}\label{eqn:logder}
    \frac{df}{dx} = \frac{-f(x + 2h) + 8f(x + h) - 8f(x - h) + f(x - 2h)}{12h},
\end{equation}
where $f$ is an arbitrary function of $x$, and $h = \Delta\ln r$ is the spacing between the sampling points. We estimate the uncertainty on the log-derivative by computing Equation \eqref{eqn:logder} for every jackknifed mean profile and taking the standard deviation of the resulting distribution. An extra multiplicative factor of $\sqrt{N - 1}$ is applied to convert the measured uncertainty to the unbiased uncertainty, and this is analogous to the extra $N - 1$ factor used in the variance estimator, as shown in Equation \eqref{eqn:Jackknife_Var}.

\textbf{Covariance of the log-derivative:} To compute a detection significance for any feature, we require the bin-to-bin covariance matrix, $\mathcal{C}$, of the measured mean log-derivative, as is discussed further below in Equation \eqref{eqn:chi2_significance}. This covariance is estimated using a jackknife sampling of the profiles,
\begin{equation} \label{eqn:Jackknife_Covar}
    C_{ij} = \frac{N(r) - 1}{N(r)}\sum_{k = 1}^{N_{\rm cl}}\,\,\bigg(f^\prime_{k, i} - \langle f^\prime\rangle_i\bigg)\bigg(f^\prime_{k, j} - \langle f^\prime\rangle_j\bigg)\delta_j,
\end{equation}
where $i$ and $j$ index over the different radial bins, $f^\prime_{k, i}$ is the log-derivative of the mean profile in the $i^{\rm th}$ bin for the $k^{\rm th}$ jackknifed sample. All quantities in the sum are implicit functions of radius, and we have suppressed the notation for brevity. The \textit{correlation} matrix is shown in Figure \ref{fig:CorrMat}.

\textbf{Quantifying feature location:} We are interested in the location of a given feature --- particularly, of local minima in the log-derivative --- and this is estimated by fitting cubic splines to the log-derivative of each mean profile in the jackknifed sample and then locating the feature of interest in each profile. The mean and standard deviation of the resulting distribution provide estimates of the location of the feature and the associated uncertainty. Given our use of the jackknife method to estimate the uncertainty, the $\sqrt{N - 1}$ factor is needed once again to convert from the measured uncertainty to the unbiased uncertainty. For the SZ-selected samples, the median uncertainty in $\Mtwohm$ (as determined from the catalogs) is $15\%$, and so the uncertainty in $\Rtwohm$ is around $5\%$. This is tolerable as it increases the total uncertainty
in the estimated feature location by $<2\%$. Note that the uncertainty in the feature location comes from variations in the \textit{shape} of the profile. This depends both on the raw signal-to-noise of the measurement \textit{and} on the intrinsic shape of the profiles. Thus, profiles that appear noisy can still have precise feature locations if the shape of the profile has less variation.

\subsection{Modeling and Detection Quantification} \label{sec:Model_detection_significance}

As was done in \citetalias{Anbajagane2022Shocks}, we look for features in the profile outskirts by comparing the measurements with theoretical predictions. The model we employ here for the halo-$y$ correlation follows that used in \citetalias{Anbajagane2022Shocks} with some changes that we highlight. 

The model consists of two components: a one-halo term given by the projected version of the pressure profile from \citet{Battaglia2012PressureProfiles}, who calibrated the profiles using hydrodynamical simulations, and a two-halo term which accounts for contributions from nearby halos as described in \citet{Vikram2017GalaxyGroupstSZ} and later in \citet{Pandey2019GalaxytSZ}. The two-halo term prediction uses a linear matter power spectrum and linear halo bias, and assumes higher-order corrections are not required. We have validated this assumption in \citetalias{Anbajagane2022Shocks} checking the model matches the two-halo term of profiles from \textsc{The Three Hundred} simulations \citep{Cui2018The300, Cui2022The300SIMBA}. The entire model is implemented in the \textsc{Core Cosmology Library (CCL)} open-source python package\footnote{\url{https://github.com/LSSTDESC/CCL}} \citep{Chisari2019CCL} and is public.\footnote{\url{https://github.com/DhayaaAnbajagane/tSZ_Profiles}}

We begin by representing the 3D, halo-pressure cross-correlation function as a composition of the one-halo and two-halo components,
\begin{equation} \label{eqn:TotHaloDecomposition}
    \xi_{h,p}(r , M, z) = \xi_{h,p}^{\rm one-halo}(r , M, z) + \xi_{h,p}^{\rm two-halo}(r , M, z),
\end{equation}
where $\xi$ are the correlation functions, $r$ is \textit{comoving} distance, and $M$ is the halo mass. We denote the combined one-halo and two-halo term as the ``total halo model''. The one-halo term is obtained via the pressure profile of \citet{Battaglia2012PressureProfiles},
\begin{equation} \label{eqn:Battaglia_Prof}
    P(x) = P_{\rm 200c}P_0\bigg(\frac{x}{x_c}\bigg)^\gamma\bigg[1 + \bigg(\frac{x}{x_c}\bigg)^\alpha\bigg]^{-\beta},
\end{equation}
where $P_0$, $x_c$, $\alpha$, $\beta$, and $\gamma$ are the fit parameters calibrated from hydrodynamical simulations, $x = r/\Rtwohc$ is the distance in units of cluster radius, and $P_{\rm 200c}$ is the thermal pressure expectation from self-similar evolution,
\begin{equation} \label{eqn:SelfSimPressureScale}
    P_{\rm 200c} = 200\rho_c(z)\frac{\Omega_{\rm b}}{\Omega_{\rm m}}\frac{G\Mtwohc}{2\Rtwohc}.
\end{equation}
Equation \eqref{eqn:Battaglia_Prof} accounts for deviations from self-similar evolution via the calibrated mass and redshift dependencies of the parameters $P_0$, $x_c$, and $\beta$. The model also includes the effects of non-thermal pressure support within halos --- which is generated by the incomplete thermalization of gas --- as it is calibrated on simulations that include this phenomenon. The fit parameters for Equation \eqref{eqn:Battaglia_Prof} are obtained from the ``200 AGN'' calibration model of \citet[see Table 1]{Battaglia2012PressureProfiles}, and these parameters have a known, calibrated scaling with both cluster redshift, $z$, and cluster mass, $\Mtwohc$. The calibration matches simulations within $<10\%$ in the one-halo regime \citep[][see their Figure 2 and Section 4.2]{Battaglia2012PressureProfiles}. While \citetalias{Anbajagane2022Shocks} used the ``500 SH'' model, the ``200 AGN'' model opted for here provides a better fit to the measured profiles on small-scales and is the model choice for other works that we compare to below (\eg in Section \ref{sec:GroupScales}). The pressure deficit we discuss below is observed regardless of the model chosen to be the comparison point.

The tSZ emission is connected to the \textit{electron} pressure, $P_e$, whereas the profiles of \citet{Battaglia2012PressureProfiles} are calibrated to the total gas pressure, $P$. We convert between them as
\begin{equation} \label{eqn:Electron_Pressure}
    P_e(r , M, z) = \frac{4 -2Y}{8 - 5Y}P(r , M, z),
\end{equation}
with $Y = 0.24$ being the primordial helium mass fraction. This provides our one-halo term,
\begin{equation} \label{eqn:OneHalo}
    \xi_{h,p}^{\rm one-halo}(r , M, z) = P_e(r , M, z).
\end{equation}

It is more convenient to compute the two-halo term in Fourier space, so our computations are done in the same. We inverse Fourier transform the model in the end to obtain the required real-space correlation function. The two-halo term of the halo-pressure cross-power spectrum, $P^{\rm two-halo}_{h,p}$, is written as,
\begin{equation} \label{eqn:PowerSpectrum}
    \begin{split}
        P^{\rm two-halo}_{h,p}(k , M, z) & = \bigg[b(M, z) \,P_{\rm lin}(k , z) \,\,\times \\
            &\qquad \int_0^\infty \kern-1em dM^\prime \, \frac{dn}{dM^\prime} \, b(M^\prime, z) \,u_p(k,M^\prime, z)\,\bigg],
    \end{split}
\end{equation}
where $M$ is the mass of the halo we are computing the halo-pressure correlation for, $M^\prime$ is the mass of a neighbouring halo contributing to the two-halo term, $P_{\rm lin}(k , z)$ is the linear, matter density power spectrum at redshift $z$, $dn/dM^\prime$ is the mass function of neighbouring halos, and $b(M , z)$ and $b(M^\prime , z)$ are the linear bias factors for the target halo and neighboring halos, respectively. The mass function model comes from \citet{Tinker2008HMF} and the linear halo bias model from \citet{Tinker2010HaloBias}. The term $u_p(k,M^\prime, z)$ is the Fourier transform of the pressure profile about the neighboring halo which, under the assumption of spherical symmetry, is computed as,
\begin{equation}\label{eqn:Battalia_Fourier}
    u_p(k,M^\prime, z) = \int_0^\infty dr 4\pi r^2 \frac{\sin(kr)}{kr}P_e(r,M^\prime, z),
\end{equation}
where $P_e$ is the electron pressure profile. The halo-pressure two-point cross-correlation is obtained as the inverse Fourier transform of the cross-power spectrum,
\begin{equation} \label{eqn:TwoHalo}
    \xi_{h,p}^{\rm two-halo}(r , M, z) = \int_0^\infty \frac{dk}{2\pi^2}k^2\frac{\sin(kr)}{kr} P_{h,p}^{\rm two-halo}(k , M, z).
\end{equation}
The terms shown in equations \eqref{eqn:OneHalo} and \eqref{eqn:TwoHalo} can be combined according to Equation \eqref{eqn:TotHaloDecomposition} to get the total halo model, $\xi_{h,p}$.

We have thus far described the real-space 3D pressure, whereas the Compton-$y$ parameter is the \textit{integrated} (or projected) pressure along the line of sight. The halo-$y$ correlation is therefore obtained by a projection integral,
\begin{equation} \label{eqn:ProjTotalHalo}
    \xi_{h, y}(r,M, z) = \frac{\sigma_T}{m_e c^2}\int_{-\infty}^{\infty}\frac{d\chi}{1 + z}\xi_{h, p}\bigg(\sqrt{\chi^2 + r^2},M, z\bigg),
\end{equation}
where $\sigma_T$ is the Thomson scattering cross-section, $m_ec^2$ is the rest mass energy of the electron, and $\chi$ is the comoving coordinate along the line-of-sight. 

All SZ maps have a finite angular resolution, where the resolution limitation suppresses power on small scales. We incorporate this into our model by smoothing the prediction. We first calculate the angular cross-power spectrum, using the flat sky approximation, as,
\begin{equation} \label{eqn:C_ellConvert}
    C_\ell = \int  \,d\theta\, 2\pi \theta\, J_0(\ell \theta)\, \xi_{h, y}(\theta ,M, z),
\end{equation}
where $J_0$ is the zeroth-order Bessel function. We then multiply $C_\ell$ by the Fourier-space smoothing function for the given survey of interest and then perform an inverse-harmonic transform,
\begin{equation} \label{eqn:TotalHaloSmoothed}
    \xi^{\rm smooth}_{h, y}(\theta , M) = \int \frac{d\ell\ell}{2\pi}J_0(\ell \theta)C_\ell B_\ell,
\end{equation}
with the smoothing function $B_\ell$ given as
\begin{equation}\label{eqn:Beam_Smoothing}
    B_\ell = \exp\bigg[-\frac{1}{2}\ell(\ell + 1)\sigma_{\rm FWHM}^2\bigg],
\end{equation}
where $\sigma_{\rm FWHM} = \theta_{\rm FWHM}/\sqrt{8\ln 2}$, with $\theta_{\rm FWHM} = 1.25^\prime$ ($\theta_{\rm FWHM} = 1.6^\prime$) being the full-width half-max of the Gaussian filter used to smooth the SPT (ACT) maps. 

Our final theory curve for a given cluster sample is obtained as follows: we compute the smoothed total halo model, $\xi^{\rm smooth}_{h, y}$, for each individual cluster in our catalog, and then perform a weighted stack identical to that done on the data, i.e. where the weights are the SNR of the observed clusters. The only inputs to this model are the cluster mass, redshift, and SNR (which is used as a weight). Thus, the theoretical curves shown below are true predictions and are not model fits made on the profile measurements. The one exception is the model for DES clusters, which includes a miscentering component (described in Section  \ref{sec:MiscenteringModel}) which \textit{does} have free parameters that we vary. The approach to fixing those parameters is described in that same section. We generally only discuss results for DES clusters that do not require a theoretical model.

Finally, we estimate the significance of any deviation between the measured log-derivatives and the theoretical model as
\begin{equation} \label{eqn:Detection_significance}
    \epsilon \equiv \frac{1}{\sigma}\bigg(\frac{\dln y^{\rm obs}}{\dln x} - \frac{\dln y^{\rm th}}{\dln x}\bigg),
\end{equation}
where $\sigma$ is the uncertainty in the log-derivative measurement. The quantity $\epsilon$ is the number of sigma by which the log-derivative in the data differs from that of the theory.

We also measure a standard chi-squared significance for the feature of interest as a whole,
\begin{equation} \label{eqn:chi2_significance}
    \chi^2 = \bigg(\frac{\dln y^{\rm obs}}{\dln x} - \frac{\dln y^{\rm th}}{\dln x}\bigg)^T \mathcal{C}^{-1} \bigg(\frac{\dln y^{\rm obs}}{\dln x} - \frac{\dln y^{\rm th}}{\dln x}\bigg),
\end{equation}
where $\mathcal{C}^{-1}$ is the inverse of covariance matrix for the log-derivative, accounting for the Hartlap factor \citep{Hartlap2007} as, 
\begin{equation} \label{eqn:Hartlap}
    \mathcal{C}^{-1} \rightarrow \frac{N_{\rm jk} - N_{\rm bin} - 2}{N_{\rm jk} - 1} \,\mathcal{C}^{-1}.
\end{equation}
where $N_{\rm jk}$ are the number of jackknife samples (more than 500 for almost all samples), $N_{\rm bins} = 5$ are the number of bins used to estimate the significance of a particular feature (i.e. the pressure deficit). The rescaling accounts for the bias due to limited realizations being used to numerically estimate the covariance matrix.
The covariance $\mathcal{C}$ is defined in Equation \eqref{eqn:Jackknife_Covar}. As mentioned above, we do not use all 50 radial bins for this calculation and instead limit ourselves to all bins whose radii are within $\Delta \log_{10}r = 0.1$ of the location of the feature. Once the $\chi^2$ is computed, we quote the total signal-to-noise of a feature, as
\begin{equation}\label{eqn:chi2sh}
    \chi_{\rm sh} = \sqrt{\chi^2 - N_{\rm dof}}
\end{equation}
following the definition of \citet[][see their Equation C15]{Secco2022MassAp}, with $N_{\rm dof} = 5$ as mentioned above. This definition of signal-to-noise improves on that used in \citetalias{Anbajagane2022Shocks} as it is more robust to noise fluctuations and binning choices.

\subsubsection{Miscentering model for optically selected clusters}\label{sec:MiscenteringModel}

An additional component to our theoretical model, in comparison to that of \citetalias{Anbajagane2022Shocks}, is the impact of cluster miscentering. For SZ-selected clusters, the offset between the cluster center and the true center (called ``miscentering'') is negligible when compared to the radial scale of features we study, which are $\sim \Rtwohm$. When using optically selected clusters, however, the optically determined center can be significantly offset from the center of the gas distribution \citep{Sehgal2013ACTMiscenter, Zhang2019MiscenteringDESY1, Bleem2020SPT-ECS}.\footnote{SZ-selected clusters also incur a noise-induced miscentering effect, with a scale of $\theta_{\rm miscen} = \sqrt{\theta^2_{\rm 500c} + \theta^2_{\rm beam}}/{\texttt{SNR}}$. For $R \gtrsim \Rtwohm$, the miscentering scale is at/below the bin width and is negligible as our features of interest span multiple bins. The average SZ-selected cluster ($\Mtwohm \approx 10^{14.8} \msol$ and $z \approx 0.6$) has $\theta_{\rm miscen} = 0.3\arcmin$, while the same for the average optically selected cluster ($\Mtwohm \approx 10^{14.6} \msol$ and $z \approx 0.4$) is factors of 5 to 10 larger  \citep{Zhang2019MiscenteringDESY1, Bleem2020SPT-ECS}.} The impact of miscentering in the profile is to transfer power from small scales to large scales. The total observed profile, with miscentering, can be modelled as,
\begin{equation}\label{eqn:MiscenCombined}
    y(R) = (1 - f_{\rm miscen})y^{\rm true}(R) + f_{\rm miscen}y^{\rm miscen}(R),
\end{equation}
where $f_{\rm miscen}$ is the fraction of miscentered objects and $y^{\rm true}$ ($y^{\rm miscen}$) is the profile of correctly centered (miscentered) clusters. For a given miscentering offset, $R_{\rm mis}$, the average miscentered profile is,
\begin{equation}\label{eqn:MiscenModelAng}
    y^{\rm miscen}(R \,|\, R_{\rm mis}) = \int_0^{2\pi} d\theta y^{\rm true}\bigg(\sqrt{R^2 + R_{\rm mis} + 2\cos\theta R R_{\rm mis}}\bigg),
\end{equation}
and the total model is obtained by marginalizing over the distribution of possible offsets,
\begin{equation}\label{eqn:TotalMiscen}
    y^{\rm miscen}(R) = \int dR P(R_{\rm mis}) y^{\rm miscen}(R \,|\, R_{\rm mis}).
\end{equation}
Following previous works \citep[\eg][]{ Baxter2017SplashbackSDSS, Chang2018SplashbackDES, Shin2019SplashbackDESxACTxSPT, Shin2021SplashbackDESxACT}, we assume the offsets follow a Rayleigh distribution,
\begin{equation}\label{eqn:OffsetProb}
    P(R_{\rm mis}) = \frac{R_{\rm mis}}{\sigma_R^2}\exp\bigg[-\frac{R_{\rm mis}^2}{2\sigma^2_R}\bigg],
\end{equation}
\begin{equation}\label{eqn:OffsetSigma}
    \sigma_R = \tau_{\rm miscen} \bigg(\frac{\lambda}{100}\bigg)^{0.2} \mpc.
\end{equation}
where $\lambda$ is the cluster richness. The free parameters of this model are $f_{\rm miscen}$ and $\tau_{\rm miscen}$ which set the fraction of miscentered objects, and the amplitude of the miscentering offset, respectively. The impact of miscentering --- and the choice of the parameter values --- for DES cluster profile model is discussed in Section \ref{sec:fiducialresults} and further in Appendix \ref{appx:Miscentering}.

\section{Shocks in Galaxy Clusters} \label{sec:Results}

We first present our main results in Section \ref{sec:fiducialresults} using the cluster samples of the different surveys, then study the variation of the profiles (i) with cluster selection and choice of SZ map in Section \ref{sec:SurveyComparisons}, and; (ii) with halo mass, towards group-scale halos, in Section \ref{sec:GroupScales}. We will use the format \texttt{CATALOG} x \texttt{MAP} as a shorthand reference for measurements for a given cluster catalog using a given SZ map (\eg SPTxSPT, DESxACT).

All bands show $68\%$ uncertainties estimated via jackknife resampling of the profiles. As for the detection significance, we show $\epsilon$ in the figures but quote $\chi_{\rm sh}$ in our discussions in the text as the total signal-to-noise of a feature. These are defined in equations \eqref{eqn:Detection_significance} and \eqref{eqn:chi2_significance}, respectively. The latter is the combined significance of the feature across multiple radial bins, while the former is the single-bin significance and is useful for identifying the radial range of a signal. 

Constraints on feature locations and their corresponding detection significance are provided in Table \ref{tab:Results}. In general, the measured location of the feature is expected to be offset from the true location due to the impact of beam smoothing in the SZ maps. However, we have verified previously, using simulations, that this difference is negligible for the SPT and ACT resolution level \citepalias{Anbajagane2022Shocks}. Note that, for the average cluster in our samples, the scale of $\Rtwohm$ is a factor of $\approx 5$ larger than the full-width half-max of the smoothing scale in these maps.

While the specific focus of this work is on finding pressure deficits and other shock-induced features in the SZ profile outskirts, this focus also requires we discuss profile behaviors in the one-halo and two-halo regimes. Shocks occur at the transition between the bound halo component (one-halo term) and the surrounding large-scale structure (two-halo term), so studying shock-induced features also requires studying these regimes. Thus, some of our discussions below will include behaviors of the one-halo and two-halo terms, as changes in these terms affect the overall shape of the halo profile.

\subsection{Measurements from fiducial cluster samples}\label{sec:fiducialresults}

\begin{figure*}
    \centering
    \includegraphics[width = 2\columnwidth]{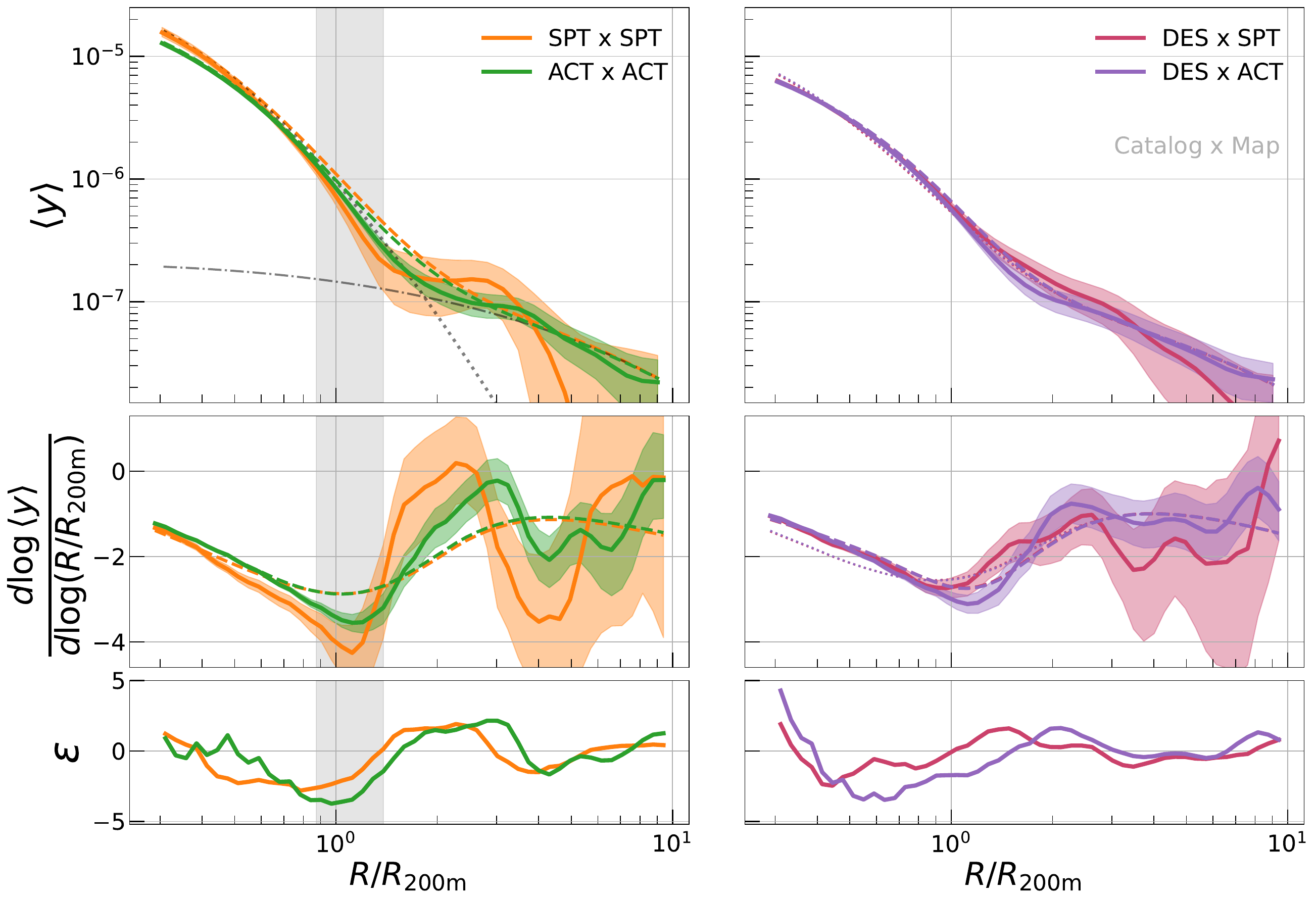}
    \caption{The average SZ profiles of different cluster samples, and measured using different SZ maps (top), their associated log-derivative (middle), and the difference between the log-derivatives of the data and model (bottom) as defined in Equation \eqref{eqn:Detection_significance}. The theoretical prediction (dashed lines), which is a sum of one-halo and two-halo contributions (each shown as gray dotted and dashed-dotted lines, respectively, in the left panel for the SPT predictions alone), is described in Section \ref{sec:Model_detection_significance}. The left panels show results for SZ-selected clusters from SPT and ACT, while the right is optically selected clusters in DES with a mass cut $\Mtwohm > 10^{14.5} \msol$. For the SZ-selected samples, the derivative is lower at $\Rtwohm$ than the theory curve, consistent with \citetalias{Anbajagane2022Shocks}. The behavior for optically selected clusters is less clear due to potential inaccuracies in the theoretical model, such as the miscentering model.  All profile measurements have a two-halo component, seen most prominently  at large radii, that is consistent with the model. Estimates for the location and depth of the first log-derivative minimum (``pressure deficit'') in each measurement are shown in Table \ref{tab:Results}. The gray band in the left panels demarcates the range of radii used to quantify the significance of the pressure deficit as shown in the table. The dotted lines in the right panel are the theory models without any miscentering effects included; including the miscentering (dashed line) changes/improves the model. The two dashed lines in the right panels overlap with one another. The correlation matrix of the log-derivative is shown in Figure \ref{fig:CorrMat}.}
    \label{fig:Profiles}
\end{figure*}

In Figure \ref{fig:Profiles}, we present the average SZ profiles of different cluster samples measured using different SZ maps. The SPT result is from the exact same data as \citetalias{Anbajagane2022Shocks}, but analyzed using the slightly updated measurement pipeline described in Section \ref{sec:Measurement}. As was the case in \citetalias{Anbajagane2022Shocks}, the theoretical prediction matches the measurements in the cluster core ($R/\Rtwohm \lesssim 0.5$) and also in the far outskirts ($R/\Rtwohm \gtrsim 5$), but has significant deviations at $R/\Rtwohm \approx 1$, and potentially also at $R/\Rtwohm \approx 3$. These two deviations were denoted a pressure deficit and accretion shock, respectively, in \citetalias{Anbajagane2022Shocks} and we use the same nomenclature here.

This pressure deficit was discussed in \citetalias{Anbajagane2022Shocks} as a possible sign of thermal non-equilibrium between electrons and ions, where the non-equilibrium is generically caused by shock heating \citep{Fox1997ElectronNE, Ettori1998ElectronNE, Wong2009ElectronNE, Rudd2009ElectronNE, Akahori2010ElectronNE, Avestruz2015ElectronNE, Vink2015ElectronNE}. Shocks are the primary mechanism for converting kinetic energy to thermal energy during structure formation. They preferentially heat the ions over the electrons given the former are more massive. Thus, shock-heated plasma has colder electrons than protons, and the low density of particles in the cluster outskirts implies these two particle species never equilibrate. \citet[see their Figure 2]{Rudd2009ElectronNE} use simulations specialized to model the electron-ion temperature differences and show that this effect causes a deficit in the cluster tSZ profiles\footnote{Such a deficit should also be present in electron temperature profiles measured through X-ray data. However, our current X-ray observations do not extend to such large radii, $R \approx \Rtwohm$, and are instead limited to much smaller radii where the higher number densities allow the ion and electrons to quickly achieve temperature equilibrium.}, while \citet[see their Figure 1]{Avestruz2015ElectronNE} do the same but focus on the 3D cluster temperature profiles. This pressure deficit feature would not be present in most cosmological hydrodynamical simulations as they a priori assume local thermal equilibrium between electrons and ions. We will henceforth refer to the pressure deficit as a shock feature and denote its location the shock radius, $\Rshock$.

As Figure \ref{fig:Profiles} and Table \ref{tab:Results} show, the ACT DR6 data strengthen the evidence for a pressure deficit feature near the cluster virial radius. This is the same feature first noted in \citetalias{Anbajagane2022Shocks} with SPT-SZ data and with ACT DR5 clusters measured on the ACT DR4 map. We estimate the significance of the feature in the ACT data at $6.1\sigma$. Given the new, more robust definition of signal-to-noise in Equation \eqref{eqn:chi2sh} and the switch from the ``Shock heating'' model of \citet{Battaglia2012PressureProfiles} to the  ``200 AGN'' model, the estimated significance of the feature in SPT-SZ is $2.7\sigma$ compared to the estimate of $3.1\sigma$ from \citetalias{Anbajagane2022Shocks}. We have verified that our pipeline reproduces the SPT-SZ result of the previous work if we revert back to the previous signal-to-noise definition and model choice.

The deficit in both SPT and ACT is found at consistent radial locations, with $\Rad = 1.09 \pm 0.08$ and $\Rad = 1.16 \pm 0.04$, respectively. The minima in the log-derivatives are consistent as well, with $\logder = -4 \pm 0.5$ and $\logder = -3.5 \pm 0.1$, respectively. These estimates are detailed further in Table \ref{tab:Results}. The similarity of the deficit seen in SPT and ACT suggests the feature is physical and not an artifact introduced in either the map-making or the cluster-finding procedures in each survey. We have also independently verified the consistency of these features using a complementary fitting method, described in Appendix \ref{appx:SamFits}. In \citetalias{Anbajagane2022Shocks}, we validated that the theoretical model used in this work matches cosmological hydrodynamical simulations (see their Figure 4). In specific, we used the \textsc{The300} suite which simulates a sizable number of massive clusters, and provides a sample relevant for SZ-selected cluster catalogs which have $\Mtwohm > 10^{14.5} \msol$. Thus, any differences between the measurements and the theoretical profiles can be equivalently interpreted as differences between the measurements and simulations. 

The bottom panels of Figure \ref{fig:Profiles} also present the quantity $\epsilon$, defined in Equation \eqref{eqn:Detection_significance}, which is the bin-by-bin deviation between the measured and predicted log-derivatives, normalized by the measurement uncertainty. In SZ-selected clusters, $\epsilon$ takes a maximum value at $\Rad \approx 1$, corresponding to the pressure deficit. In optically selected clusters, which we will discuss below, the maximum values of $\epsilon$ are at smaller scales. This is because the measurement is much more precise on these scales so small deviations between the data and theory --- such as those caused by imperfections in the miscentering model ---  can have large statistical significance.

We do not discuss the potential accretion shock features in detail as these are currently still low-significance features dominated by noise, as was the case in \citetalias{Anbajagane2022Shocks}. We simply note it is intriguing that the log-derivatives of the SPT and ACT profile measurements both have a maximum at $\Rad \approx 3$, followed by a sharp drop. The maximum corresponds to a plateauing phase in the profiles, which is a feature of the accretion shock as presented in \citet{Baxter2021ShocksSZ}. More detailed work is required to robustly verify this feature as arising from the presence of a shock.

Other studies also find features in the cluster outskirts using a variety of different datasets. \citet{Hurier2019ShocksSZPlanck} see a sharp decrease in pressure at $R = 3\Rfivehc \approx \Rtwohm$ for a single cluster in the \textit{Planck} data. \citet{Pratt2021ShocksPlanck} also use \textit{Planck} data and find an \textit{excess} in pressure at $R = 2\Rfivehc \approx 0.7\Rtwohm$ for a set of ten, low-redshift galaxy groups. The analysis of \citet{Planck2013PressureProfiles} finds that the 3D pressure profiles have a deficit, relative to the theoretical predictions of \citet{Battaglia2012PressureProfiles}, for $R \gtrsim \Rfivehc$ ($R \gtrsim 0.3\Rtwohm$) while being a good match for scales below that radius. \citet{Zhu2021ShockXray} find an excess in the temperature and density profiles of the Perseus cluster at $R \approx \Rtwohc = 0.5\Rtwohm$ using \textit{Suzaku} X-ray data. \citet{Hou2023ShockRadio} study the radio emission around galaxy clusters and find a signal at $R = 2.5\Rfivehc \approx \Rtwohm$. They interpret this as the presence of a non-thermal electron population and find that the corresponding electron energy distribution is consistent with one generated by strong shocks.  In all works, the deviations are found around $R \approx \Rtwohm$, consistent with the shock radius $\Rshock$.

The right panels of Figure \ref{fig:Profiles} show, for the first time, the outskirts of SZ profiles for \textit{optically selected} clusters. We have placed a mass cut of $\Mtwohm > 10^{14.5} \msol$ (where $\Mtwohm$ is the mass inferred from the cluster richness, see Section \ref{sec:DES_Data}) on this sample as this reduces the impact of systematic effects (such as projection, contamination, etc.); this cut is also consistent with the minimum mass of the SZ-selected samples (see Figure \ref{fig:Characterize_Dataset}). We discuss the results of lower mass objects, which are removed by this cut, in Section \ref{sec:GroupScales}. The SZ profiles of DES clusters have a $\approx 30\%$ lower normalization than those of the SZ-selected catalogs, and this is due to the differences in mass distributions and the mean mass of the samples (see Table \ref{tab:Results}). The normalization of the theoretical model (dashed lines) also decreases a similar amount if we input the DES cluster mass/redshift distribution rather than the SPT or ACT ones.  At $\Rtwohm$, which is inbetween the one-halo and two-halo regime, the profile for the DESxACT measurement has a minimum log-derivative ($\logder = -3.1 \pm 0.15$) that is more negative than that of the DESxSPT measurement ($\logder = -2.7 \pm 0.2$), with a significance of  $1.6\sigma$. The two results use different cluster subsamples, defined as all DES clusters within the ACT/SPT footprint. We interpret this difference as a statistical variation and do not examine it further. We verify in Figure \ref{fig:MapCheck} below that the SPT and ACT maps provide statistically indistinguishable results across the full range of scales considered in this work.

Looking at the DESxACT and the ACTxACT results, we see the location of the log-derivative minima is consistent at $0.2\sigma$, while the depth of minima is deeper in ACTxACT at $2\sigma$. The comparison of DESxSPT and SPTxSPT is similar, where the location of the log-derivative minima is consistent while the depth deviates at $2.4\sigma$. The mass and redshift distributions of the DES cluster sample are notably different from those of ACT and SPT, which could lead to differences in this depth. In Section \ref{sec:SurveyComparisons}, we re-analyze the ACT and DES data after accounting for such mass/redshift differences, and find that the depth becomes consistent across the two measurements.

The model (dashed line) for the DES-related results in Figure \ref{fig:Profiles} is a qualitatively good match to the data across the whole range of presented scales. The prediction for the DESxSPT and DESxACT measurements closely overlap one another. This model includes the miscentering effects described in Section \ref{sec:MiscenteringModel}, using values of $\tau_{\rm miscen} = 0.9$ and $f_{\rm miscen} = 0.4$. These values were chosen after exploring a sparsely sampled 2D grid of parameter values and picking the parameters that provided the visually best fit to the one-halo regime, near the cluster core. The preferred values for $\tau_{\rm miscen}$ and $f_{\rm miscen}$ are both near the $3-4\sigma$ upper limit of the miscentering parameter constraints of \citet[][see their Chandra--DES constraints in Table 1]{Zhang2019MiscenteringDESY1} for the DES Y1 cluster sample. However, the value of $\tau_{\rm miscen}$ is within $1\sigma$ of the estimate from \citet[][see their Table 6]{Bleem2020SPT-ECS}, which is based on a SPT-DES matched cluster sample. Figure  \ref{fig:Profiles} shows the theory matches the data better (in the 1-halo regime) when we include this miscentering effect, and the dotted lines show the theory without such effects.

In Appendix \ref{appx:Miscentering}, we discuss how the profiles and log-derivatives depend on miscentering parameters. We emphasize that in our work we only focus on results from optically selected clusters that are insensitive to the choice of miscentering model and parameters. For example, Table \ref{tab:Results} does not quote any detection significance of a pressure deficit for DES clusters. However, we still measure and quote the location and depth of the log-derivative minimum for the DES cluster profiles as it does not depend on an assumed theoretical model.

Our results show that the SZ-selected clusters have a clear pressure deficit while such a deficit is not seen as clearly in optically selected clusters. In general, this difference could occur if (i) SZ-selected clusters have a selection effect that preferentially picks out objects with such features, (ii) an aspect of the richness--SZ--mass correlations makes optically selected clusters suppress the deficit feature, and (iii) systematic effect(s) in optically selected clusters (\eg the miscentering, contamination, or mass estimation errors) causes the feature to be suppressed. In Section \ref{sec:SurveyComparisons} below, we verify that the first two possibilities are not the cause for the difference between the results of SZ-selected and optically selected clusters. The third possibility --- the systematic effects in optically selected clusters --- is an intricate issue spanning many different parts of the cluster detection/processing pipeline, and so we do not explore this direction as it is beyond the scope of our work. However, in Section \ref{sec:SurveyComparisons}, we will show that limiting the DES clusters to higher masses, $\langle \Mtwohm \rangle = 10^{14.85} \msol$, results in the profile measurement showing a deficit that is consistent with those of the SPT and ACT clusters. This in turn implies that the three effects mentioned above have negligible impact on the measurements if we use optically selected clusters that are limited to higher masses than those of the fiducial sample used in Figure \ref{fig:Profiles}.

One SZ-related systematic effect is the CIB, which is sourced by dusty, star-forming galaxies. DES clusters are selected on richness (i.e. galaxy counts) and preferentially contain clusters with more satellite galaxies compared to an SZ-selected sample. Thus, the amplitude of the infrared signal for such a sample could be higher. However, SZ-selected samples probe higher redshifts than optically selected clusters, which are closer to the peak of cosmic star-formation at $z = 2$. We verify in Appendix \ref{sec:CIB} that our results are unchanged if we use SZ maps that minimize/deproject the CIB signal.

\subsection{Sensitivity to map-making and cluster selection}\label{sec:SurveyComparisons}

\begin{figure}
    \centering
    \includegraphics[width = \columnwidth]{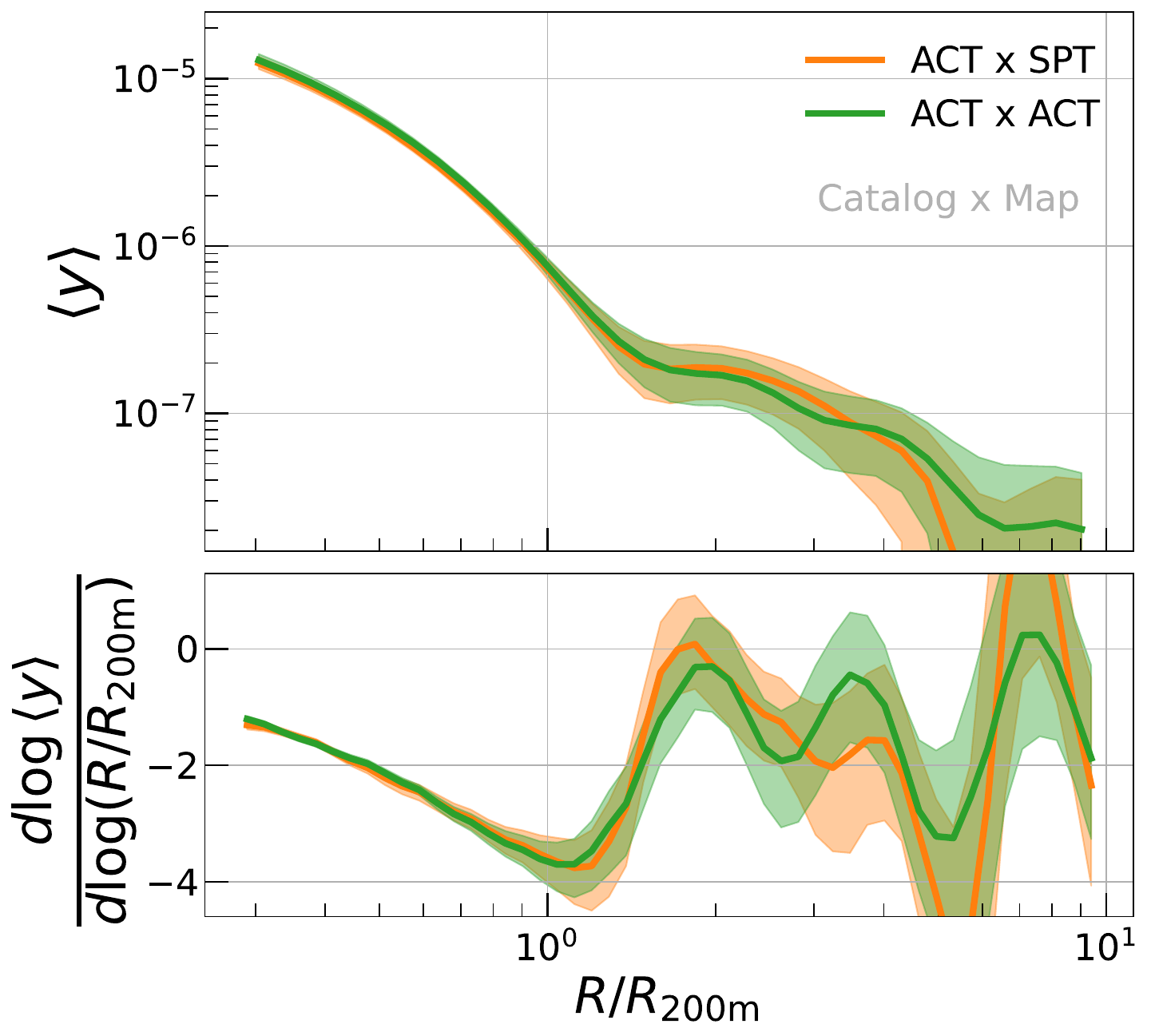}
    \caption{The average SZ profile of an ACT cluster subsample ($N = 669$ clusters) measured using either the ACT map or SPT map. The subsample is defined as all clusters whose centers lie in both the ACT and SPT footprints. The two measurements are consistent across the whole range of scales, with $\chi^2/N_{\rm dof} = 1.1$ and $p = 0.14$, validating that the datasets and map-making procedures of the two surveys are consistent in both high and low signal-to-noise regimes. The SPT and ACT datasets are independently calibrated and mapped, and the statistical consistency in the measurements above is determined at the $\approx 1\%$ level given the precise measurements in the high signal-to-noise regime.}
    \label{fig:MapCheck}
\end{figure}

\begin{figure}
    \centering
    \includegraphics[width = \columnwidth]{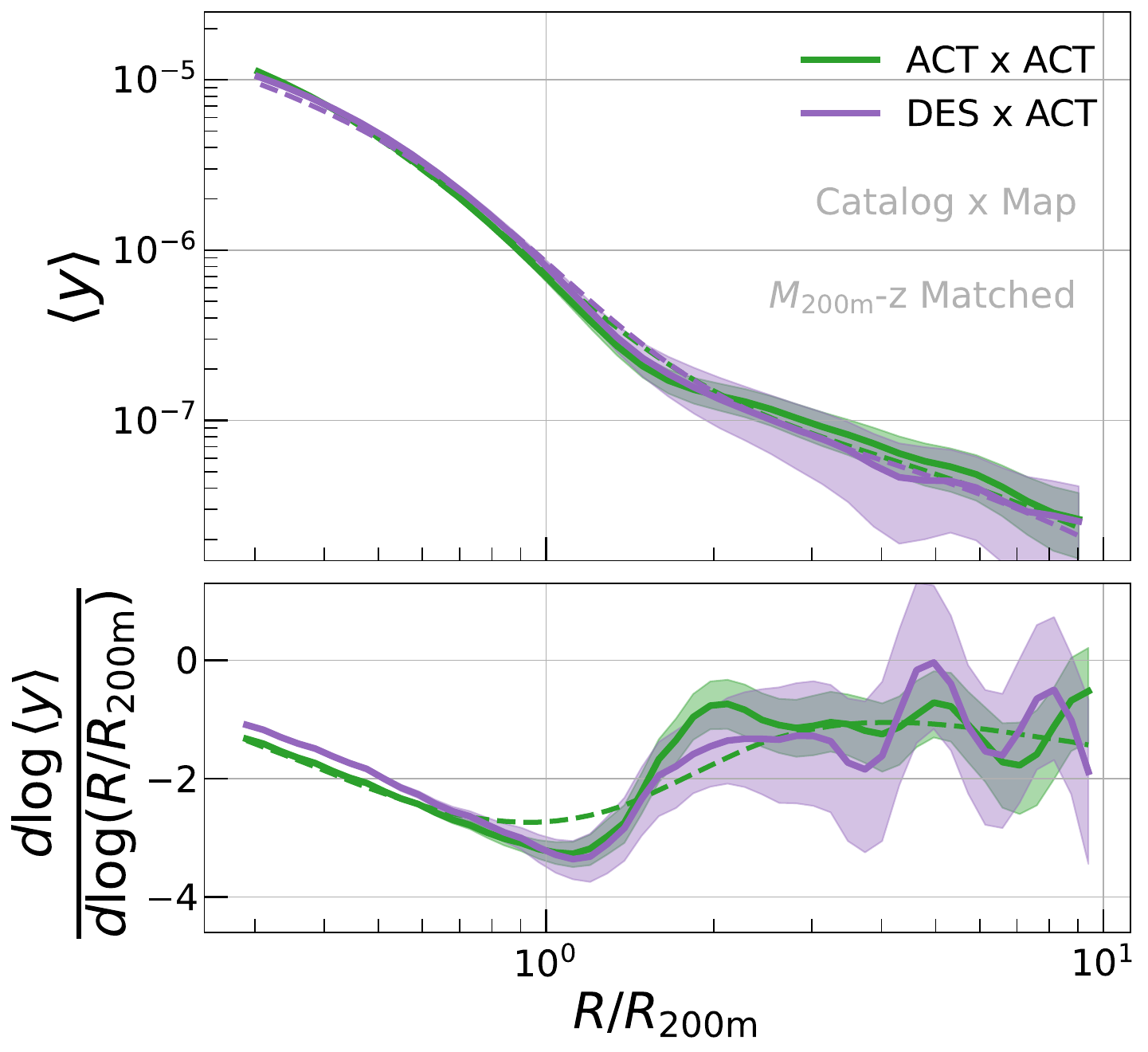}
    \caption{The average SZ profiles of two different cluster samples: an SZ-selected one from ACT, and an optically selected one from DES. In both cases, the samples are modified from the original distribution. We use all ACT clusters with $0.1 < z < 0.8$, and then reweight the DES clusters to match the ACT subsample's $\Mtwohm - z$ distribution. See Section \ref{sec:SurveyComparisons} for details. The two profiles are consistent ($\chi^2/N_{\rm dof} = 1.1$ with $p = 0.14$), suggesting that in this mass/redshift range there are no SZ or optical selection effects that generate/suppress the pressure deficit feature.}
    \label{fig:ClusterSelection}
\end{figure}

The results in Figure \ref{fig:Profiles} show a difference between the profiles of SZ-selected and optically selected clusters --- the former sees a clear pressure deficit at $R/\Rtwohm \approx 1$, while the latter either sees a less significant feature or no feature at all --- and this could be caused by SZ-selection preferentially picking out clusters with such deficit-like features, or by the optical selection effects preferentially missing such clusters (in this case, due to a correlation this feature may have with cluster richness). 

In Figure \ref{fig:MapCheck}, we test an aspect of the former effect, namely noise-based SZ-selection effects.\footnote{We consider this a systematics-based selection effect, in contrast to physical selection effects such as, for example, SZ-selected clusters being preferentially more/less dynamically active compared to mass-selected clusters. An aspect of these physical SZ-selection effects is tested in Figure \ref{fig:ClusterSelection}.} These effects correspond to the fact that the clusters are identified in the same (noisy) maps used to measure their SZ profiles. We test the impact of this effect by taking all ACT clusters that fall into the intersection of the ACT and SPT footprints ($N = 669$ clusters), and then by measuring the subsample's average SZ profile using either the SPT map or the ACT map. We find consistency ($\chi^2/N_{\rm dof} = 1.1$ with $p = 0.14$) in both the profiles and the log-derivatives of the two measurements.  While this implies that noise-based SZ-selection effects are not the cause of the pressure deficit feature, the agreement is also a check on the data and map-making procedures of the SPT and ACT surveys.\footnote{A similar analysis using all SPT clusters in both footprints finds $\chi^2/N_{\rm dof} = 1.06$ with $p = 0.36$. However, the profile measurement uncertainties are broader as the SPT cluster sample size is half that of ACT.} It validates the maps' consistency in both the high signal-to-noise regime at the location of massive clusters, as well as in the noise-dominated, low-signal-to-noise regime of the cluster outskirts. 

Next, we test the impact of optical selection on this deficit feature by comparing profiles around ACT and DES clusters that are reweighted to have the same mass/redshift distribution. The reweighting is done to minimize any differences in the measured average SZ profiles due to differences in just the mass/redshift distribution of the samples.\footnote{The zeroth-order effect of SZ and optical selection on the cluster sample is in its mass and redshift distributions (see Figure \ref{fig:Characterize_Dataset}). The reweighting accounts for these selection effects, and thus any further differences in the reweighted profiles can be attributed to selection effects beyond those on the cluster samples' mass and redshift distributions.} We first remove all ACT clusters with redshifts/masses outside the ranges of the DES sample. We therefore use all ACT clusters within $0.1 < z < 0.8$ and $13.7 < \log_{10}\Mtwohm < 15.35$ to create the subsample used in this analysis, which has $N = 3392$ clusters. The reweighting is then done by computing the weighted counts of clusters in a 2D grid of $\Mtwohm$ and $z$, and then using the ratio of ACT counts to DES counts. The weight used in the weighted counts is the signal-to-noise per cluster, consistent with the rest of our analysis. The exact expression of the re-weighting is
\begin{equation}\label{eqn:reweights}
    w(\Mtwohm, z) = \sum^{\rm N_{\rm cl, ACT}}_{\rm i = 1} \delta_i \texttt{SNR}^{\rm ACT}_i {\bigg/} \sum^{\rm N_{\rm cl, DES}}_{\rm i = 1} \delta_i \texttt{SNR}^{\rm DES}_i,
\end{equation}
where $\delta_i$ is a delta function --- with values of 0 or 1 --- that denotes whether cluster $i$ falls into a given mass and redshift bin. We compute the weights in a 10-by-10 grid and assign each DES cluster a new weight based on the $\Mtwohm-z$ grid cell it is associated with. We have checked that our results do not change if we use a 20x20 grid instead. The final weight of the DES cluster is,
\begin{equation}\label{eqn:new_weight}
  w^{\rm mod}(\Mtwohm, z) = \texttt{SNR} \times w(\Mtwohm, z),
\end{equation}
which uses the original signal-to-noise weights of our analysis alongside the mass/redshift-based reweighting of Equation \eqref{eqn:reweights}. We have tested that our results, shown below, are unchanged if we exclude all ACT clusters in the DES footprint, where this exclusion would remove any overlap between the cluster samples. 

Figure \ref{fig:ClusterSelection} shows the average SZ profile around the ACT subsample and the reweighted DES sample. The two profiles are consistent with one another. The ACT subsample shows a clear pressure deficit in the log-derivatives --- evidenced by the measured profile dropping more steeply at $R/\Rtwohm \approx 1$ than the theoretical prediction --- and the DES measurement matches this feature. This consistency is partially expected as the DES reweighting increases the contribution of the most massive clusters to the average SZ profile, and any systematic effects on the pressure deficit measurement could be less prominent in this mass regime. However, it is still a valuable check as even in the high-mass regime, optically selected clusters have shown differences in their total matter density profiles that were due to optical selection effects \citep{Baxter2017SplashbackSDSS, Chang2018SplashbackDES, Shin2019SplashbackDESxACTxSPT}.

Figure \ref{fig:ClusterSelection} provides evidence that at high mass, the optical selection does not result in biased SZ profiles for the one-halo and two-halo regimes. Across the range $0.5 < R/\Rtwohm < 10$, the two profiles are consistent($\chi^2/N_{\rm dof} = 1.4$ with $p = 0.1$). Under this reweighting, the weighted mean mass of the DES sample increases from $\langle\Mtwohm\rangle = 10^{14.6}\msol \rightarrow 10^{14.85}\msol$, which is a fractional change of 80\%, while the mean redshift is left unchanged at $\langle z \rangle = 0.46$ (see Table \ref{sec:Results}). The depth of the minima is now consistent across the two samples, whereas it was inconsistent at the $2\sigma$ level for the fiducial ACT and DES cluster samples (Figure \ref{fig:Profiles}). Given that agreement between the samples is recovered after accounting for their mass/redshift differences, we infer that the earlier disagreement was due to these differences.

The results of Figure \ref{fig:MapCheck} and Figure \ref{fig:ClusterSelection} imply that --- for a mass and redshift range corresponding to clusters in SZ surveys (see Figure \ref{fig:Characterize_Dataset}) --- the SZ or optical selection has negligible impact on the measured pressure deficit. This adds to the robustness of the deficit features found in the SPTxSPT and ACTxACT measurements, as clusters identified with a completely different type of data (i.e. optical images) still show a pressure deficit. These results also show that the DES sample exhibits a clear deficit (given its agreement with the ACT measurement) when limited to higher masses, implying that the shallower log-derivative depth found in our fiducial measurement (Figure \ref{fig:Profiles}) could possibly be attributed to the clusters in the lower mass end of the sample. We explore the behavior of such systems further in the following section.


\subsection{Towards galaxy groups}\label{sec:GroupScales}

\begin{figure*}
    \centering
    \includegraphics[width = 2\columnwidth]{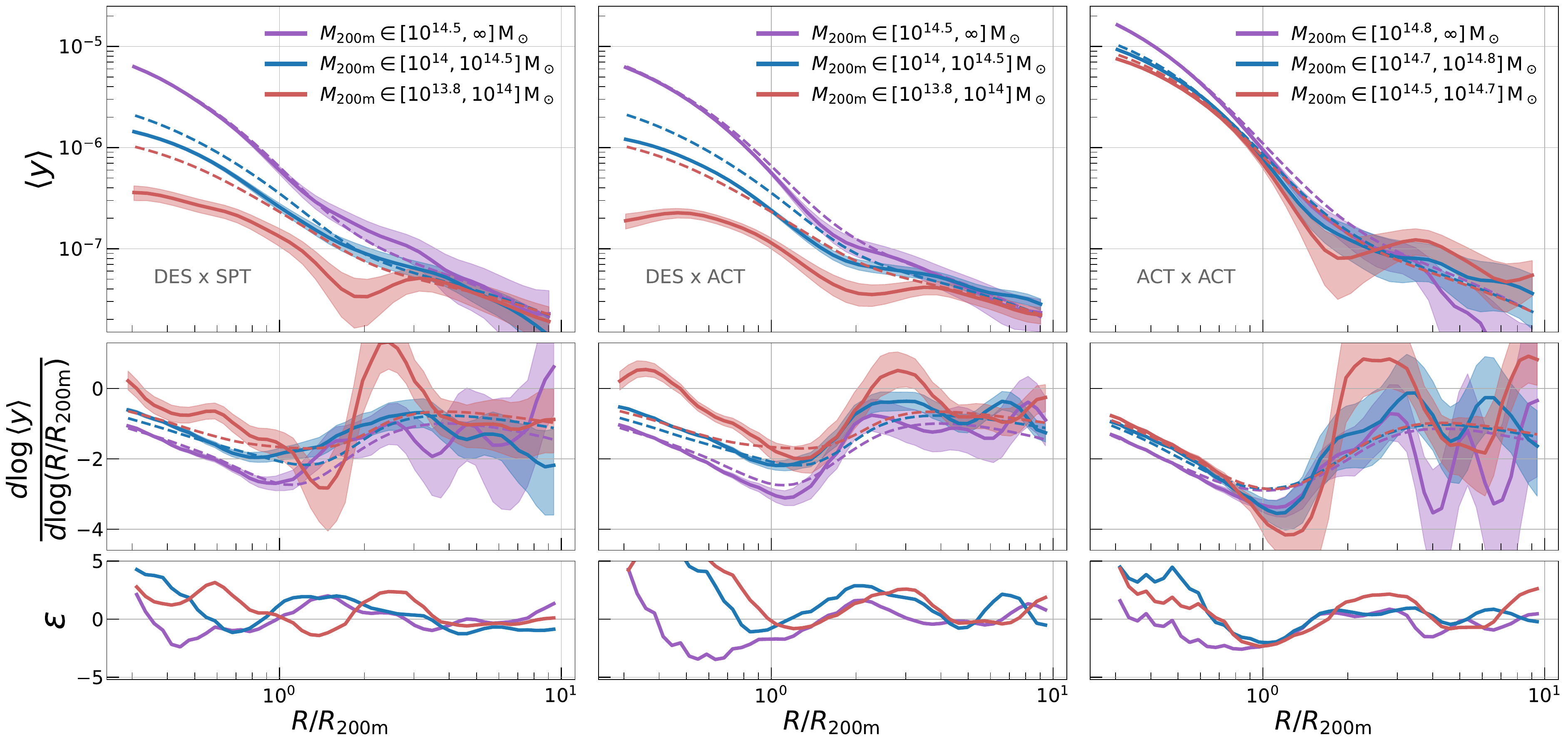}
    \caption{The average SZ profiles, binned according to inferred halo mass, for different combinations of cluster samples and SZ maps. The dashed lines are the theoretical prediction described in Section \ref{sec:Model_detection_significance}. Note that the mass ranges for the ACT sample (right) are significantly narrower than those of the DES sample (left, middle). The measured profiles of lower-mass clusters ($\Mtwohm < 10^{14.5} \msol$) deviate significantly from the theoretical predictions, but the two-halo term is still consistent between data and theory. This two-halo term is prominent at large radii, as shown in Figure \ref{fig:Profiles}.}
    \label{fig:GroupScales}
\end{figure*}

The profiles of massive clusters have many observational constraints, especially near the cluster core \citep[$R/\Rtwohm \lesssim 0.5$, see for example,][]{McDonald2014SPTGasProfiles, Ghirardini2017ChandraPressureProfile, Romero2017MUSTANGPressure, Romero2018MultineProfile, Ghirardini2018ShockAbellXMM}, and the further outskirts have only been recently explored observationally \citep{Planck2013PressureProfiles, Sayers2013SZBolocam, Sayers2016BolocamPlanck, Amodeo2021ACTxBOSS, Schaan2021ACTxBOSS, Melin2023SZProfiles, Lyskova2023Xray3R500c}. Less massive objects --- galaxy group scales and down to Milky Way scales --- have not been studied on a profile level, and the presence/absence of any features in the outskirts is relatively unknown. Previous works have studied the cross-correlation function of the tSZ field with galaxy counts \citep{Hill2018tSZxGroups, Amodeo2021ACTxBOSS,  Schaan2021ACTxBOSS, Sanchez2023DESxSPT}, which is an observable that is sensitive to halo profiles but cannot always distinguish features in the profiles. For example, the pressure deficit in the cluster outskirts (Figure \ref{fig:Profiles}) was not identified in previous cross-correlation works but was easily identified in \citetalias{Anbajagane2022Shocks} by measuring individual profiles. SZ-selected halo samples are ideal for studying massive, cluster-scale halos but are not viable for probing lower masses. Here, we use the DES \textsc{redMaPPer} sample to obtain a catalog of lower mass objects ($\Mtwohm \gtrsim 10^{13.8} \msun$) and measure their SZ profiles across a wide range of scales.

In Figure \ref{fig:GroupScales} we show the average SZ profile for three mass bins of DES clusters, measured on both the SPT map (left) and the ACT map (middle), and also the profiles for the ACT cluster sample measured on the ACT map (right). The mass bins of the latter differs significantly from those of the former two. Focusing first on the ACT cluster results in the rightmost column of Figure \ref{fig:GroupScales}, we see the pressure deficit exists for all mass bins, at $3.8 \sigma$, $2.5\sigma$, and $2.9\sigma$ from highest to lowest mass bins. The three minima from the log-derivative measurements are all statistically consistent with each other. This result also serves as an additional validation check --- if the pressure deficit in SZ-selected samples is caused by noise-based selection effects, then its amplitude will be higher for clusters detected at the low signal-to-noise regime, which is right near the cluster detection threshold. In Figure \ref{fig:GroupScales}, however, we find that splitting by mass --- which is directly proportional to signal-to-noise --- does not notably change the significance of the deficit. In Appendix \ref{appx:SNRDependence},  we also redo this test by splitting directly on SNR instead of $\Mtwohm$, and find consistent results.

The other two columns in Figure \ref{fig:GroupScales} (left and middle) show the SZ profile for three mass bins of optically selected clusters. The mass range $\Mtwohm > 10^{14.5} \msol$ corresponds to $\lambda > 30$, while the range $10^{14}\msol < \Mtwohm < 10^{14.5} \msol$ corresponds to $15 < \lambda < 30$, and finally, $10^{13.8} \msol < \Mtwohm < 10^{14}\msol$ corresponds to $10 < \lambda < 15$. These masses are not exact translations of the richness but rather approximate conversions for interpreting the discussions to follow. The highest mass bin (purple) is the same result as Figure \ref{fig:Profiles} and shows good, qualitative agreement between the measurements and the theoretical predictions. When comparing the measurements of lower mass bins to those of the highest mass bin, we see that for lower mass objects the log-derivatives are closer to zero in the one-halo term ($\Rad \lesssim 1$) and similar to the high mass bin results for the two-halo term ($\Rad \gtrsim 4$). The log-derivative minima in each mass bin are found at similar radii of $\Rad \approx 1.1$ (see Table \ref{tab:Results}).

The theoretical model also significantly deviates from the measurements in these two lower mass bins. For $\Mtwohm \in [10^{13.8}, 10^{14}]\msol$, the deviation is a factor of $\approx 5$ in the halo core. In the intermediate mass bin, $\Mtwohm \in [10^{14}, 10^{14.5}]\msol$, it is a factor of $\approx 2$ and is also consistent with previous analyses in this intermediate mass range. \citet[][see their Table 1 and Figures 5/6]{Saro2017SZoptical} found a factor of $\approx 2$ difference when measuring the integrated SZ effect around clusters from the DES Science Verification data, while \citet[][see their Figure 2]{Planck2011SuppressSZ} finds similar suppression in the SZ-richness scaling relation. These differences generically point to some inaccuracy in the theoretical model. 

The discussion of the mass trends thus far focuses on the behavior of the log-derivative minima, rather than of the  ``pressure deficit''. The latter is defined as significant deviations between measurement and model in the shape of the profile. However, for the lower mass bins, the model has inaccuracies as noted above, which limit our ability to identify such a feature. There are a few known reasons why such inaccuracies could occur: (i) the contamination of the cluster sample at low masses, \eg two or more low-mass clusters are projected together on the sky and are observed as one large cluster, causing a mass estimation bias (ii) inaccuracy in the utilized pressure profile model for lower mass halos, and; (iii) significant correlations in the richness and SZ scatter at fixed halo mass which, in tandem with the optical selection effect at low richness, could become an important effect. We briefly discuss each to check if it can explain the deviations and thereby provide an avenue to correct the existing model prediction.

\textbf{The first, contamination of the sample}, causes an overestimate of the cluster mass (and thus, the SZ profile) compared to the truth. This overestimate is more significant in the one-halo regime than the two-halo regime as the latter's mass-dependence is weaker. The two-halo term scales as $y \propto b_h(M) \propto M^{0.5}$ for the halo bias model of \citet{Tinker2010HaloBias} at high halo masses, whereas the one-halo term scales as $y \propto M^{5/3}$. The deviations in Figure \ref{fig:GroupScales} are roughly factors of 2-5 in the SZ signal, and suggest the corresponding maximum bias in the mass --- assuming a self-similar scaling of $y \propto \Mtwohm ^ {5/3}$ --- would be $50\%$ to $150\%$ in $\Mtwohm$. \citet{Myles2021Projection} show the richness bias due to contamination is $\approx 20\%$ for clusters of $5 < \lambda < 20$ (see their Section 4.3), and also that richness depends on halo mass as $M \propto \lambda ^ {1.0}$ (see their Section 4.5). This implies the mass bias is  $20\% \times 1.0 = 20\%$, lower than the required values of 50\% to 150\% denoted above, and provides evidence that contamination from projection cannot be the dominant cause of the suppression. Similarly, variations in the assumed projection model of the mass--richness relation show $\approx 30\%$ changes in the final mass estimate \citep[][see their Equations 16 and 17]{Costanzi2021DESxSPT}.

\textbf{The second effect, $Y-M$ relation deviations}, are deviations in the pressure profile model for lower mass halos. This work uses the model of \citet{Battaglia2012PressureProfiles}, and while it is accurate for higher mass halos (\eg see Figure \ref{fig:Profiles}), observational analyses find a preference for deviations from this model at lower halo masses \citep{Hill2018tSZxGroups, Pandey2021DESxACT}. Such deviations can arise from differences between the assumed galaxy formation process in the simulations, and the relevant processes in the data. In particular, these above works suggest the SZ signal for the lower mass bins we consider here is suppressed by factors of 3-4 and that the suppression grows stronger with decreases in halo mass. Both these behaviors are consistent with our findings. However, the uncertainties on the inferred suppression are not precise enough to confirm that this effect is the dominant cause of the deviations in Figure \ref{fig:GroupScales}.

\textbf{The third effect, correlations in the richness and SZ scatter at fixed mass}, is relevant as our work involves the simultaneous use of cluster mass, SZ, and richness; we select clusters using richness, infer a halo mass from this richness, and then use the inferred halo mass to predict the SZ profile. The correlations between the three properties require non-trivial corrections to the model for the SZ--mass scaling relation of the selected cluster sample. The effect has been detailed in the analytical work of \citet[][see their Figure 4 for an example]{Evrard2014}. The scaling relation for the optically selected sample is now written as
\begin{align}\label{eqn:E14Scaling}
    \langle \ln y\,|\, \ln \lambda\rangle = &\,\,  \langle \ln y\,|\, \ln\Mtwohm(\ln\lambda)\rangle \nonumber\\
    &+\,\, \beta_1 \alpha_\lambda \times {\rm cov}(\ln y, \ln\lambda | \ln\Mtwohm(\lambda)),
\end{align}
where $-\beta_1$ is the slope of the halo mass function at a chosen mass scale, $\alpha_\lambda$ is the slope of the richness--mass relation, and ${\rm cov}(\ldots)$ is the covariance of the SZ and richness scatter. A general form of this expression can be found in \citet[][see their Equation 6]{Evrard2014}. Inspecting Equation \eqref{eqn:E14Scaling} shows $\langle \ln y | \ln \lambda \rangle$ can be higher (lower) than $\langle \ln y | \ln \Mtwohm(\ln \lambda)\rangle$, for a positive (negative) sign of correlation in the SZ--richness scatter at fixed mass. \citet{Farahi2019AntiCorr} observationally constrain this correlation coefficient to be $-0.5 \lesssim r \lesssim 0.5$ at $95\%$ confidence, and their results indicate the correlation of gas-based and stellar-based cluster observables is negative (see their Table 2). Cosmological simulations also show the correlation is negative --- the scatter of gas mass and stellar mass are anti-correlated \citep[][see their Figure 5]{Farahi2018Bahamas} while that of the stellar mass and richness are correlated \citep[][see their Figure 7]{Anbajagane2020StellarProperty}. A negative correlation/covariance suppresses the SZ signal of richness-selected clusters, which could cause the observed suppression. For conservative values of $-\beta_1 = 1.5$, $\alpha_\lambda = 1.5$, $\sigma_{\ln y} = 0.3$, $\sigma_{\ln \lambda} = 0.8$, $r = -0.6$,\footnote{$\beta_1$ is the slope at a pivot mass of $\Mtwohm \sim 10^{14} \msol$, computed using the halo mass function of \citet{Tinker2008HMF}, $\alpha_\lambda$, $\sigma_{\ln y}$, and $\sigma_{\ln \lambda}$, are chosen to be larger than constraints from \citet[][see their Table 4]{Costanzi2021DESxSPT} and $r$ is set by the $95\%$ bound from \citet{Farahi2019AntiCorr}} we find the bias is at most $30\%$. Thus, this effect cannot be the main cause of the behaviors found in Figure \ref{fig:GroupScales}.

Our discussions and estimates above indicate the deviations between measurement and model are unlikely to be explained by just one of these effects. Thus, the model we use for SZ profiles cannot be easily corrected to match our measurements in the one-halo regime of lower mass clusters. Furthermore, the latter two effects we discuss --- deviations in the Y-M relation and the correlated richness and SZ scatter --- are also functions of radius that are not well-known and would be required to accurately correct our model. Previous works (including all works cited above) have only discussed these effects for volume-integrated quantities, rather than for radial profiles. Accurate predictions for these profiles, however, are necessary to study a pressure deficit (i.e. shock-induced deviations between the data and model). Given this limitation, our main results of this section focus on the raw log-derivative measurements (rather than inferring a pressure deficit from them by comparing to theory), which have a clear striation with mass in the one-halo regime and weak-to-no striation in the two-halo regime. The minima of these derivatives are located at similar radii for all three mass bins.

\section{Connections to structure formation features}\label{sec:StructureFormation}

Having explored the average SZ profiles using different combinations of cluster samples and SZ maps, we now connect these profiles to broader features from structure formation. First, we detail the connection to cosmic filaments in Section \ref{sec:Filaments} via oriented stacking of the profiles. Then in Section \ref{sec:Splashback}, we compare the pressure deficit seen in the SZ profile to the splashback feature observed in the galaxy number density profile measured around clusters.

As mentioned prior, discussing shock-induced features also requires discussing behaviors in the one-halo and two-halo regimes as shocks occur at the transition between the two. Therefore, some of our discussions below include the behaviors of these two regimes.

\subsection{Connections to filaments}\label{sec:Filaments}

\begin{figure}
    \centering
    \includegraphics[width = 0.75\columnwidth]{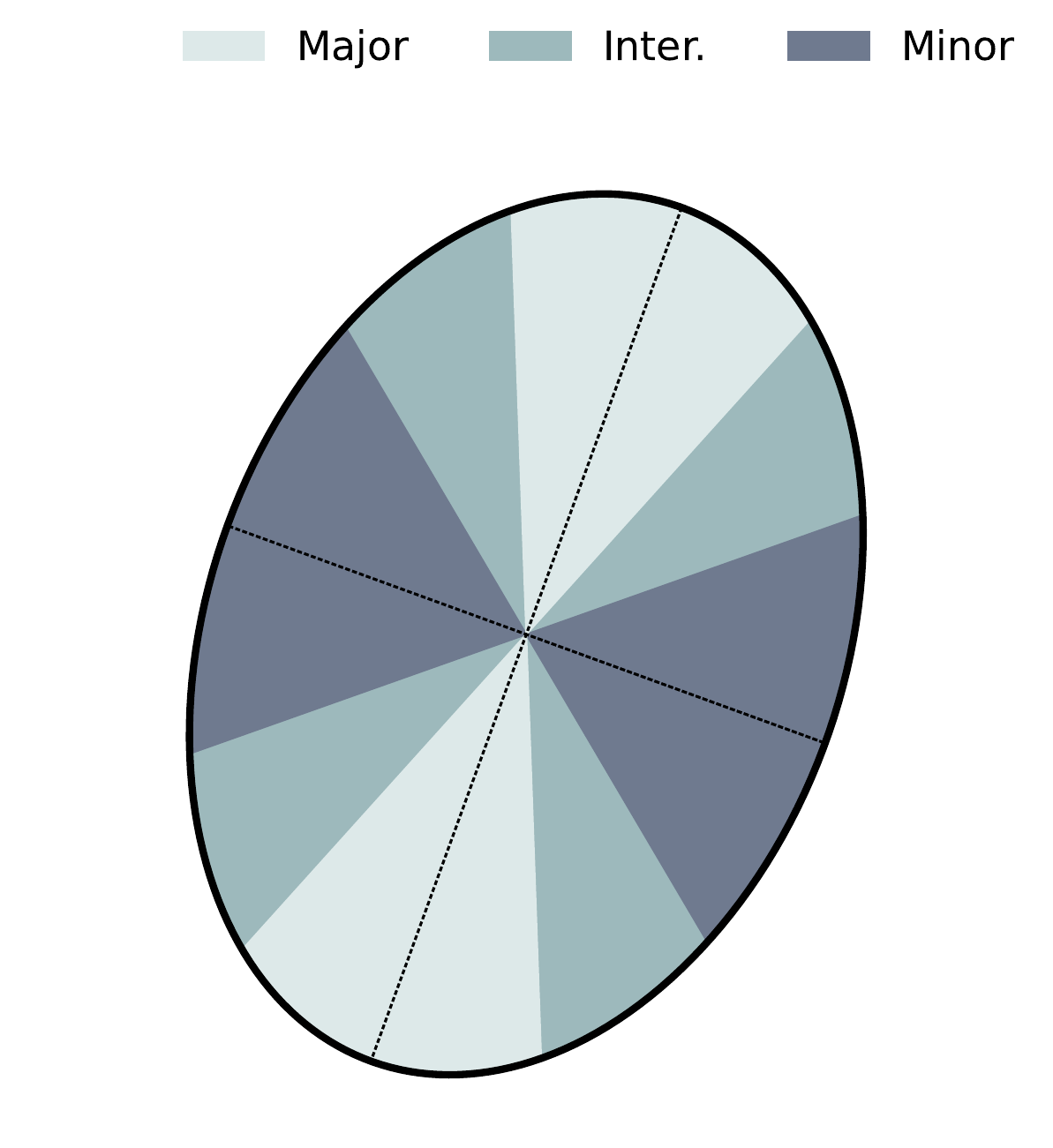}
    \caption{A diagram of how we split the cluster into three regions based on the angle away from the major axis. The lightest region falls along the major axis, the darkest along the minor axis, and we also add an intermediate region that is at $45\deg$ to both axes. Having three regions allows us to more clearly and robustly identify trends as we move from major to minor axis.}
    \label{fig:EllipticityDiagram}
\end{figure}

\begin{figure*}
    \centering
    \includegraphics[width = 2\columnwidth]{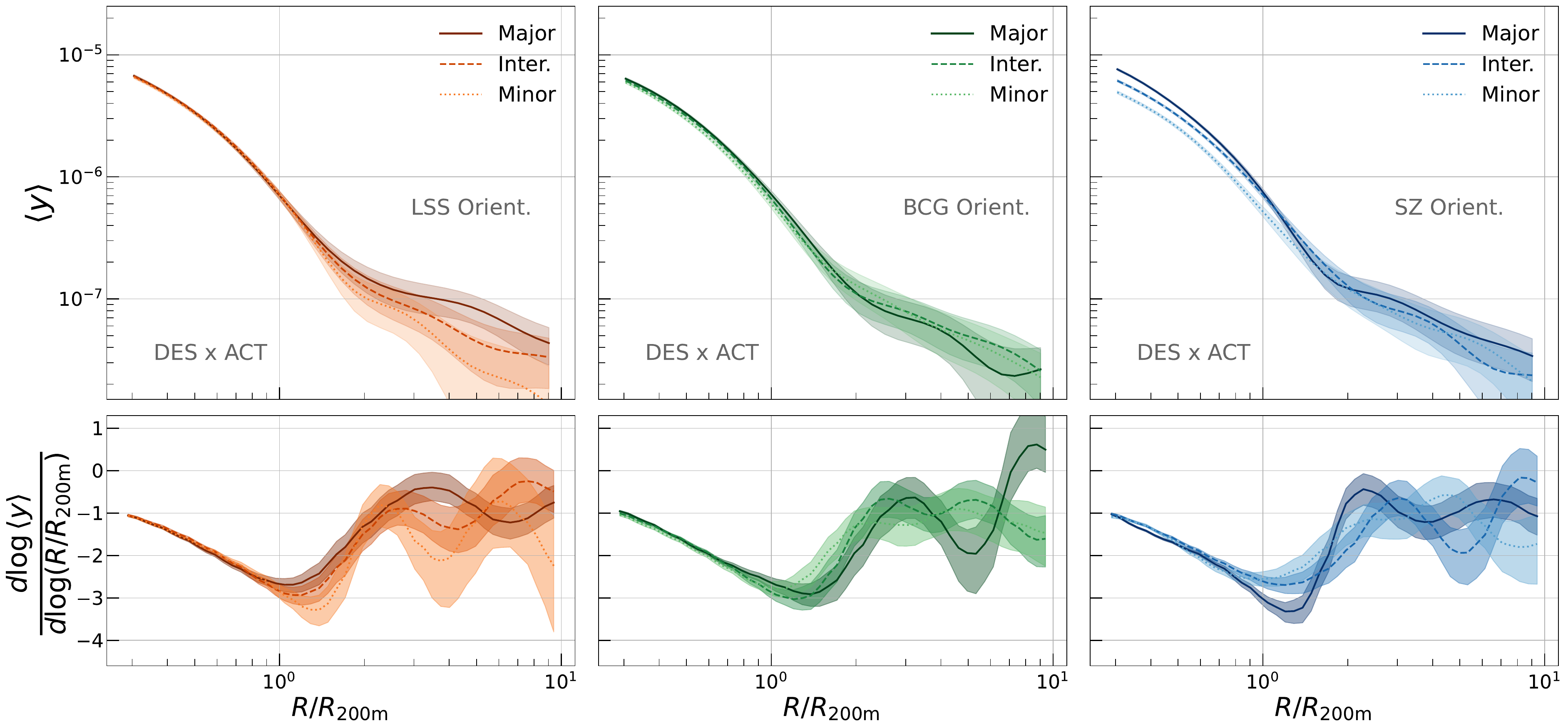}
    \caption{Oriented stacking of DES clusters on the ACT SZ map using three different methods to obtain the orientation of each cluster: the large-scale density field estimated using DES \textsc{Maglim} galaxies, the brightest cluster galaxy (BCG) shapes from the DES Y3 shape catalog, and from a 2D Gaussian fit to the SZ image cutout of each cluster. The LSS (BCG) orientation primarily impacts the two-halo (one-halo) regime of the profile. The SZ orientation (which is measured within $0.3\Rtwohm$) causes a much larger difference in the one halo term given we measure the shapes and the profiles on the same map. The BCG shape probes the orientation on scales of the order $100 \kpc$, while the SZ shape probes $0.5 \mpc$ to $1 \mpc$ scales, and the LSS-based method probes $\approx 15\mpc$ scales.}
    \label{fig:OrientationChoice}
\end{figure*}

We have discussed previously that cosmological shocks form from the accretion of collisional matter onto bound objects, and the accreted matter originates primarily from cosmic filaments. Simulations suggest that the shock boundary generally follows the same ellipticity/orientation as the cluster's, which in turn is informed by the filaments' topology around the cluster \citep[][see their Figure 1]{Aung2020SplashShock}. However, along the specific line of sight connecting the cluster core and the filament, the accretion rate of cold gas (cold relative to the hot gas bound in the cluster) is highest and can push the shock feature further into the cluster core and/or completely destroy it \citep[][see their Figure 6]{Zhang2020MergerAcceleratedShocks}.

In our analysis, we use various orientation measures, each probing a different range of scales, as estimates of the orientation of the nearby filamentary structure around the cluster. We then split the 2D SZ image of each cluster into three equal-area sections --- according to how close a section is to the major axis of the orientation --- and compute the profile using pixels within each sub-section of the image. The geometry of this split is shown in Figure \ref{fig:EllipticityDiagram}. In \citetalias{Anbajagane2022Shocks}, we split the cluster into two equal areas, corresponding to the major and minor axis. In this work, we add a third area that probes the intermediate region. Through this, we can more easily distinguish coherent trends across the orientations from any noise fluctuations. This increase in subsections is made possible by our larger cluster sample and thus, greater statistical constraining power.

We now have multiple choices for determining the orientation of the cluster. In \citetalias{Anbajagane2022Shocks}, we fit a 2D Gaussian to the SZ image and determined the cluster orientation accordingly. However, it is problematic to measure the orientation using the same data used to measure the profiles, as this can lead to a noise bias. For example, \citetalias{Anbajagane2022Shocks} limited their fits to $R < 0.5\Rtwohm$ as at larger radii the noise in the maps biased the shape measurements and the ensuing oriented profile measurements. In this work, we further alleviate this issue by only using pixels within $R < 0.3\Rtwohm$ as we do not use or show the profiles in this radial range. This can only partially, not totally, alleviate the noise bias, as the noise in the SZ map can be correlated on large scales due to the presence of the CMB and CIB contaminants. 

To make measurements that do not have such biases, we also leverage the optical survey data to obtain two completely independent estimates of the cluster orientation. In particular, we orient the clusters using the shape measurements of the brightest cluster galaxy (BCG) from the DES Y3 shape catalog, and also using the large-scale density field estimated from the distribution of DES Y3 galaxy positions. The BCG of each cluster is identified with \textsc{RedMaPPer}, and its shape is measured using the \textsc{Metacalibration} estimator. To estimate the orientation of the large-scale density field, we compute the Hessian of the smoothed, projected galaxy overdensity field. This is obtained using the methods of \citet{Lokken2022Superclustering}. As a brief description, this Hessian is a matrix of second derivatives with respect to the 2D projected coordinates, $\mathcal{H}_{ij} = \frac{\partial^2 \delta}{\partial x_i\partial x_j}$, where $\delta$ is the overdensity field of the galaxy number density and $x_i$ are the projected coordinates. The Hessian is then diagonalized to find the orientation of the major axis. In this work, $\delta$ is given by the galaxy positions of the DES Y3 \textsc{Maglim} sample \citep{Porredon2021Maglim} and is smoothed with a Gaussian filter with full-width half-max (FWHM) of $20\mpc$. In practice, this produces orientations similar to top-hat smoothing of radius $15\mpc$. On such scales, the shape measurement is dominated by the surrounding large-scale structure (i.e. filaments) and is not impacted by the cluster's own shape. More details on the method can be found in Section 3 of \citet{Lokken2022Superclustering}, and the choices used for this analysis are identical to those of that work.

Given two of the three orientation estimates come from the optical data, we focus the analysis of this section on DES clusters. For simplicity, we only show measurements made on the ACT map but note that those of the SPT map are qualitatively similar. Figure \ref{fig:OrientationChoice} shows the average SZ profiles of DES clusters measured in the three sections, where the orientation is obtained from each of the three methods listed above: the density field's Hessian, the BCG, or the SZ image. We will discuss the results of each orientation method separately.

\textbf{First, the LSS orientation.} The one-halo term of the profile is consistent across all three sections. This is expected as this method measures orientations of the density field on scales of $\approx 15\mpc$, which is well into the two-halo regime of the cluster ($R/\Rtwohm \gtrsim 5$). In the far outskirts, $R/\Rtwohm > 4$, the measured two-halo term shows a clear striation, where the amplitude grows in the direction of more structure (i.e. the major axis). In the transition regime, $R/\Rtwohm \approx 1$, the pressure profile has a steeper derivative along the minor axis. The profiles also show a plateauing feature at $R/\Rtwohm \approx 3-4$, where this plateau is found at larger radii along the major axis than the minor axis. This plateau could indicate a shock as has been shown in previous simulation work \citep{Baxter2021ShocksSZ}. If this is indeed a shock feature, its dependence on orientation would be consistent with previous work showing the shock boundary is elliptical with the major axis aligned towards the surrounding LSS from which matter is accreted \citep[][see their Figure 1]{Aung2020SplashShock}. The prevalence of this feature in all three data subsets suggests it is physical, and also adds some validity to the second minimum seen in the angle-averaged ACTxACT and SPTxSPT results in Figure \ref{fig:Profiles}. We have verified that all shock behaviors discussed above are also found when the LSS orientations are computed with a different DES Y3 galaxy sample, \textsc{redMaGiC} \citep{Rodriguez-Monroy2022RedMagic}.\footnote{The galaxies in \textsc{redMaGiC} are more conservatively selected than \textsc{MagLim}. This enables better photometric redshift precision but at the cost of a smaller sample; \textsc{redMaGiC} has approximately one-third of the galaxy counts of \textsc{MagLim}.} The consistency of the anisotropic profiles on small-scales is not an artifact of angular resolution limits. For the average cluster, the scale $R = 0.5\Rtwohm$ is a factor of $\approx 2.5$ larger than the FWHM of the smoothing for the maps. Furthermore, as we will show below, in Figure  \ref{fig:OrientationChoice}, orienting with the SZ image does show clear striations on these small-scales.

\textbf{Second, the BCG orientation.} There is now a small striation in the one-halo term, where the profile along the major axis (solid line) has a slightly higher amplitude than that along the minor axis (dotted line). The BCG in massive clusters has a size of roughly $\sim 100 \kpc$, which is a much smaller physical scale than those the other orientation estimates are sensitive to, and the direction towards large-scale structure can change noticeably across different scales \citep[see their Figure 16]{Lokken2022Superclustering}. We find that orienting by the BCG shape impacts only the one-halo regime. The observed striation in Figure \ref{fig:OrientationChoice} is the expected consequence of an elliptical cluster profile. This can be seen by taking a circular pressure profile and stretching/squeezing it to make it elliptical. At a fixed physical radius, the profile value along the minor axis will be lower (since the profile has been squeezed radially) compared to the profile value along the major axis (where the profile has been stretched radially). We do not see any clear trends in the two-halo term nor in the transition regime with the log-derivative minimum.

\textbf{Finally, the SZ orientation.} This is the technique used previously in \citetalias{Anbajagane2022Shocks}. There is a significant striation in the one-halo term that is suppressed as we move to the two-halo term. This behavior is expected as the orientation is measured on the same image used to measure the profiles. Thus the striation in the one-halo term is stronger than when using the BCG or LSS-based orientations. Note, however, that the orientation was measured using data at smaller radial scales than the lower radial limit of the profiles shown here. Near the one-to-two halo transition regime of these profiles, the log-derivative minimum along the cluster major axis (solid line) is steeper than that along the minor axis (dotted line). While this is more statistically significant than the striations seen in the LSS and BCG orientation cases, it is still not significant enough to consider a definite detection. Also, note that though the amplitude of the one-halo term varies between minor axis to major axis, the actual shape of the profile --- as seen in the log-derivatives --- is consistent in all three directions, up to a radius of $R/\Rtwohm \lesssim 0.8$. 

In summary, we observe potential behaviors of the log-derivative minima as we shift from major axis to minor axis: when orienting by the large-scale density field, the minimum along the minor axis is steeper than that along the major axis. We also see a potential sign of a shock at much larger radii; namely, the plateauing phase of the profiles. If this corresponds to an accretion shock, it implies that such oriented stacking could be a more optimal way to detect such features. This is consistent with \citet{Aung2020SplashShock}, who showed the accretion shock is elliptical and pointed along the large-scale structure, and also consistent with \citet{Baxter2021ShocksSZ}, who found the shock signal is more prominent in the azimuthally averaged profiles of relaxed clusters, as such clusters are predominantly spherical.

\subsection{Connections to splashback radius}\label{sec:Splashback}

\begin{figure}
    \centering
    \includegraphics[width = \columnwidth]{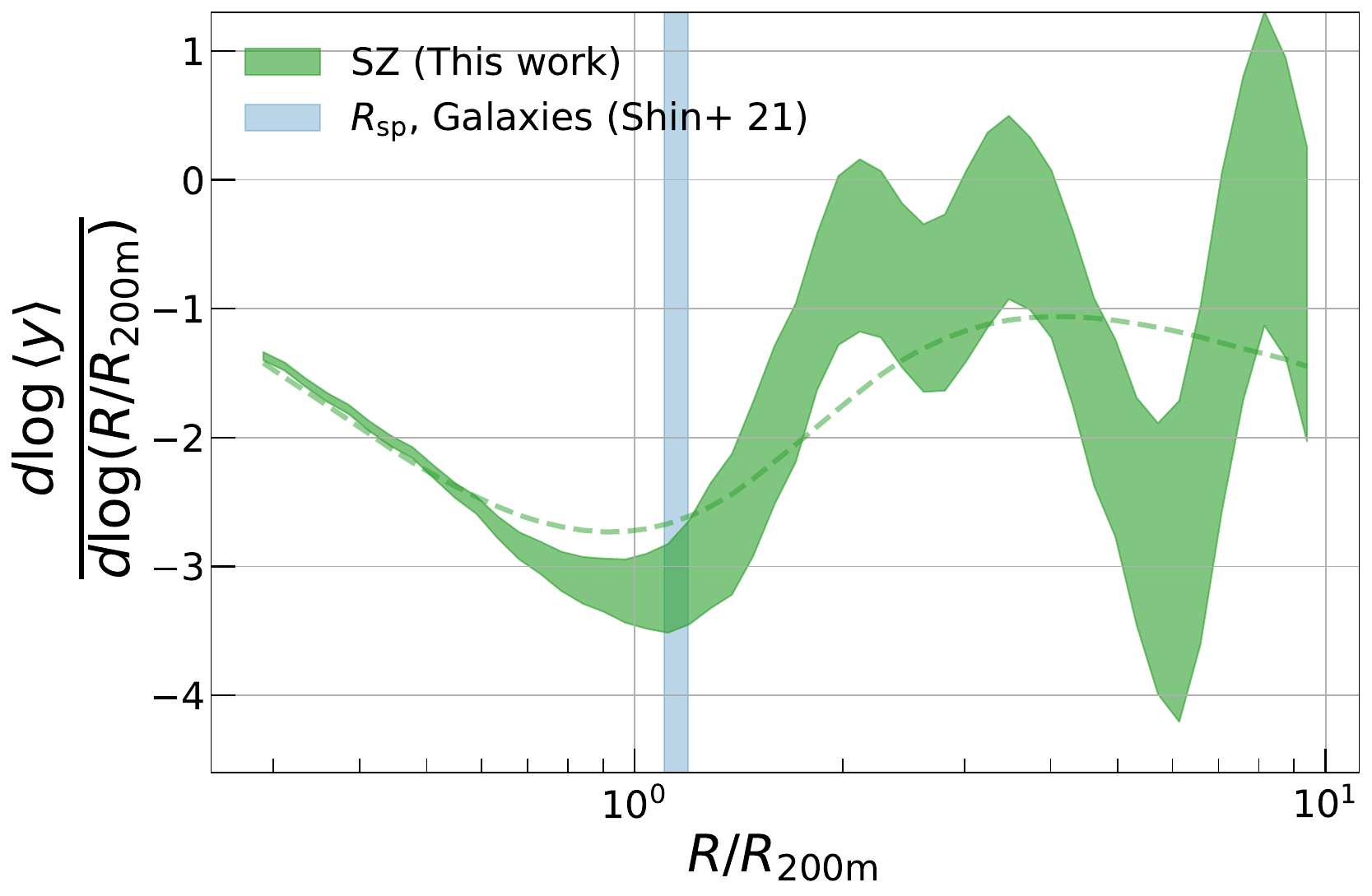}
    \caption{The log-derivative of the average SZ profile around ACT clusters with $0.15 < z < 0.7$ and whose centers are in the DES footprint. We perform this selection so as to use the same ACT DR5 subsample as \citet{Shin2021SplashbackDESxACT}, who measured the splashback radius around this sample from the galaxy number density profile (using DES Y3 galaxies) of these clusters. We show their 68\% bounds for the projected splashback radius as the vertical blue band. The minimum corresponding to the pressure deficit coincides with the splashback radius. The ratio $\Rsp/\Rshock = 1.17 \pm 0.20$, meaning the two projected radii are within $0.9\sigma$ of each other. The dashed line is the prediction of the SZ profile log-derivative for this cluster sample.}
    \label{fig:Splashback_comparison}
\end{figure}

The process of matter accretion can also cause/impact distinct features in other halo profiles, and not just the pressure profile we study here. The splashback radius, which is one such feature, is a physically motivated halo boundary defined by the apocenter in the dark matter phase space of the halo \citep[\eg][]{Diemer2014Splashback, Adhikari2014Splashback, More2015Splashback, Mansfield2017Splashback, Aung2020SplashShock, Xhakaj2020Splashback, ONeil2021SplashbackTNG, Dacunha2021SplashbackTNG}. The existence of the splashback feature has been observed by various analyses \citep{More2016SplashbackSDSS, Baxter2017SplashbackSDSS, Chang2018SplashbackDES, Shin2019SplashbackDESxACTxSPT, Zurcher2019SplashbackPlanckxPanStarrs, Murata2020SplashbackHSC, Adhikari2020SplashbackCosmicClock, Shin2021SplashbackDESxACT}, where it is identified as a minimum in the log-derivative of the lensing or galaxy number density profile --- similar to how the pressure deficit is a minimum in the log-derivative of the pressure profile --- and has been shown to play a role in galaxy formation physics \citep{Baxter2017SplashbackSDSS,Shin2019SplashbackDESxACTxSPT, Adhikari2020SplashbackCosmicClock, Dacunha2021SplashbackTNG}. The ratio of the shock radius and splashback radius, alongside appropriate theoretical models \citep[\eg][]{Shi2016ShockMAR}, can provide observational constraints on both the adiabatic index of the gas and the mass accretion rate of the cluster \citep[\eg][]{Hurier2019ShocksSZPlanck}.

\citetalias{Anbajagane2022Shocks} performed the first comparison of the splashback and pressure deficit features using the SPT-SZ dataset. In this work, we supplement this result with a complementary analysis using the ACT data. A subset of the ACT cluster catalog has already been used to identify the splashback radius, using galaxy number density profiles and weak lensing profiles measured with DES Y3 galaxies \citep{Shin2021SplashbackDESxACT}. Here, we perform the same ACT cluster catalog selections as \citet{Shin2021SplashbackDESxACT}, taking all ACT DR5 clusters within the DES Y3 footprint and with $0.1 < z < 0.7$. We then measure the log-derivative of the average SZ profile for this subsample, and present the result in Figure \ref{fig:Splashback_comparison}. We also overplot the constraint from \citet{Shin2021SplashbackDESxACT} for the splashback radius. Note that this radius was obtained by taking their fits to the observed 2D galaxy number density profile, and using the pipeline from this work to compute the minima of the log-derivative. Thus, we consistently compare the shock and splashback features in 2D, projected profiles. We do not show the weak-lensing result from \citet{Shin2021SplashbackDESxACT} but it was shown in that work to be consistent with the galaxy splashback radius.

In Figure \ref{fig:Splashback_comparison},  the minimum in the SZ log-derivative, corresponding to the pressure deficit and which we denote as the shock radius $\Rshock$, coincides with the splashback radius, $\Rsp$. In particular, we find the ratio to be
\begin{equation}\label{eqn:SplashShockResult}
    \Rsp/\Rshock = 1.17 \pm 0.20.
\end{equation}
The estimate for $\Rshock$ used above is presented in Table \ref{tab:Results} under the name ``ACT x ACT ($\Rsp$ comparison)''. Our results above show the splashback radius and shock radius are consistent within $0.9\sigma$. These results match those of \citetalias{Anbajagane2022Shocks}, who found the shock radius in SPT-SZ clusters was also consistent with the splashback radius of that same cluster sample as measured in \citet{Shin2019SplashbackDESxACTxSPT}. While some theoretical works quote a ratio of a shock radius to splashback radius \citep{Molnar2009ShocksInSZ, Shi2016ShockMAR, Aung2020SplashShock, Zhang2021SplashShock, Baxter2021ShocksSZ}, this is for the \textit{merger-accelerated accretion shock} which is expected to be at $R > \Rtwohm$ \citep{Zhang2019MergerShock}, and is distinct from the pressure deficit we discuss here. If the deficit is indeed generated by a shock, it would be linked to a typical accretion shock formed via the accretion of gas onto the halo; this is the shock we discussed in the introduction of this work and forms around the virial radius, similar to the splashback feature. Other merger-related shocks within the halo (such as bow shocks from infalling galaxies and gas clumps) can collide with this accretion shock and form a \textit{merger-accelerated accretion shock }\citep{Zhang2019MergerShock} at larger radii than the original accretion shock.

\section{Summary and Discussion} \label{sec:Discussion_Conclusions}

\begin{table*}
    \centering
    \begin{tabular}{c|l|l|c|c|c|c|c|c}
        Dataset & $\Rshock/\Rtwohm$ & $\frac{\dln y}{\dln R}(\frac{\Rshock}{\Rtwohm})$ & $\chi_{\rm sh}$ & $\langle \log_{10}\Mtwohm \rangle[\msol]$ & $\langle z \rangle$ & $N_{\rm cl}$ & Figure\\
        \hline
        SPT x SPT & $1.09\pm 0.08$ & $-3.98 \pm 0.48$ & $2.7\sigma$ & 14.94 & 0.57 & 503 & \ref{fig:Profiles}\\
        ACT x ACT & $1.16 \pm 0.04$ & $-3.53 \pm 0.12$ & $6.1\sigma$ & 14.84 & 0.55 & 4045 & \ref{fig:Profiles} \\
        DES x SPT & $0.95 \pm 0.09$ & $-2.71 \pm 0.2$ & --- & 14.63 & 0.44 & 1990 & \ref{fig:Profiles}\\
        DES x ACT & $1.14 \pm 0.07$ & $-3.11 \pm 0.15$ & --- & 14.61 & 0.44 & 4340 & \ref{fig:Profiles}\\
        \hline
        ACT x SPT (overlap) & $1.15 \pm 0.07$ & $-3.79 \pm 0.49$ & $2.6\sigma$ & 14.83 & 0.58 & 669 & \ref{fig:MapCheck}\\
        ACT x ACT (overlap) & $1.15 \pm 0.05$ & $-3.77 \pm 0.44$ & $2.7\sigma$ & 14.83 & 0.58 & 669 &\ref{fig:MapCheck}\\
        \hline
        ACT x ACT (sel. effect) & $1.11 \pm 0.06$ & $-3.26 \pm 0.18$ & $4.5 \sigma$ & 14.84 & 0.46 & 3297 & \ref{fig:ClusterSelection}\\
        DES x ACT (sel. effect) & $1.14 \pm 0.03$ & $-3.17 \pm 0.17$ & --- & 14.83 & 0.46 & 4034 &\ref{fig:ClusterSelection}\\
        \hline
        DES x SPT (high M) & $0.95 \pm 0.09$ & $-2.71 \pm 0.2$ & --- & 14.63 & 0.44 & 1990 & \ref{fig:GroupScales}\\
        DES x SPT (med M) & $0.83 \pm 0.19$ & $-1.87 \pm 0.16$ & --- & 14.19 & 0.51 & 20712 & \ref{fig:GroupScales}\\
        DES x SPT (low M) & $1.44 \pm 0.1$ & $-2.76 \pm 0.88$ & --- & 13.91 & 0.55 & 20973 & \ref{fig:GroupScales}\\[5pt]
        DES x ACT (high M) &$1.14 \pm 0.07$ & $-3.11 \pm 0.15$ & --- & 14.61 & 0.44 & 4340 & \ref{fig:GroupScales}\\
        DES x ACT (med M) & $1.10 \pm 0.12$ & $-2.23 \pm 0.14$ & --- & 14.21 & 0.50 & 45851 & \ref{fig:GroupScales}\\
        DES x ACT (low M) & $1.27 \pm 0.45$ & $-1.82 \pm 0.36$ & --- & 13.88 & 0.55 & 47426 & \ref{fig:GroupScales}\\[5pt]
        ACT x ACT (high M) & $1.13 \pm 0.08$ & $-3.39 \pm 0.19$ & $3.8\sigma$ & 14.97 & 0.51 & 1635 & \ref{fig:GroupScales}\\
        ACT x ACT (med M) & $1.15 \pm 0.08$ & $-3.56 \pm 0.33$ & $2.5\sigma$ & 14.75 & 0.57 & 1183 & \ref{fig:GroupScales}\\
        ACT x ACT (low M) & $1.23 \pm 0.15$ & $-4.15 \pm 0.68$ & $2.9\sigma$ & 14.64 & 0.62 & 1217 & \ref{fig:GroupScales}\\
        \hline
        ACT x ACT ($\Rsp$ comparison) & $1.00 \pm 0.17$ & $-3.19 \pm 0.20$ & $4.3\sigma$ & 14.86 & 0.45 & 1138 & \ref{fig:Splashback_comparison}\\
        \hline
        SPT x SPT (high SNR) & $1.11 \pm 0.04$ & $-4.27 \pm 0.66$ & $2.4\sigma$ & 15.02 & 0.56 & 259 & \ref{fig:SNRDependence}\\
        SPT x SPT (low SNR) & $0.97 \pm 0.15$ & $-3.68 \pm 0.94$ & $1.4\sigma$ & 14.81 & 0.58 & 272 & \ref{fig:SNRDependence}\\[5pt]
        ACT x ACT (high SNR) & $1.18 \pm 0.08$ & $-3.41 \pm 0.23$ & $2.5\sigma$ & 14.97 & 0.55 & 1401 & \ref{fig:SNRDependence}\\
        ACT x ACT (med SNR) & $1.19 \pm 0.05$ & $-3.65 \pm 0.38$ & $2.5\sigma$ & 14.75 & 0.57 & 1394 & \ref{fig:SNRDependence}\\
        ACT x ACT (low SNR) & $1.13 \pm 0.06$ & $-4.01 \pm 0.68$ & $4.0\sigma$ & 14.69 & 0.53 & 1400 & \ref{fig:SNRDependence}\\
        \hline
    \end{tabular}
    \caption{A summary of the numerical results presented in this work. All uncertainties are $\pm1\sigma$ estimates. From left to right the columns show: (i) the sample name,  denoted as ``cluster catalog source'' x ``SZ map source'', (ii) location of the pressure deficit, (iii) the value of the log-derivative at the location, (iv) detection significance of the feature, extracted using Equations \eqref{eqn:chi2_significance} and \eqref{eqn:chi2sh}, (v - vi) the weighted mean of the log-mass and the redshift of the cluster sample (using cluster \texttt{SNR} as weights), (vii) the number of clusters in the sample, and (viii) the Figure in this work containing the profile corresponding to the result. We do not quote a detection significance for the optically selected clusters given the dependence of this significance on the assumed miscentering model (see Section \ref{sec:fiducialresults} and Appendix \ref{appx:Miscentering} for details). The uncertainties are estimated via jackknife resampling (see Section \ref{sec:Measurement}) and do not include systematic uncertainties.}
    \label{tab:Results}
\end{table*}

The outskirts of galaxy clusters are where the collapsed halo component interacts most dynamically with the surrounding large-scale structure. A striking feature of this dynamic environment is shocks. The formation and evolution of these shocks have a rich and interesting phenomenology; they form due to the interplay between gravitational infall and hydrodynamical forces, and impact a wide array of cluster astrophysical processes once formed. In this work, we advance on previous studies and use nearly $10^5$ clusters across three datasets --- the Dark Energy Survey Year 3, the South Pole Telescope SZ survey, and the Atacama Cosmology Telescope DR4, DR5, and DR6 --- to search for shock-generated features in the average pressure profiles of different cluster samples, as measured in different SZ maps. Our key findings are summarized below:

\begin{itemize}
    \item Consistent with \citetalias{Anbajagane2022Shocks}, there is a pressure deficit at $R/\Rtwohm \approx 1.1$ detected at $2.7\sigma$ and $6.1\sigma$ in SPT and ACT, respectively (Figure \ref{fig:Profiles}). This feature is consistent with a shock-driven thermal non-equilibrium between electrons and ions. We do not quote a detection significance for DES clusters given uncertainties in the theoretical modelling (see Section \ref{sec:fiducialresults}).\vspace{0.5em}
    
    \item The SZ maps from SPT and ACT are consistent in both high and low signal-to-noise regimes (Figure \ref{fig:MapCheck}). For a subset of clusters that lie within both SPT and ACT footprints, we measure the mean SZ profiles using either the SPT map or ACT map and show the profiles are consistent across the entire radial range of our analysis, $0.3 < R/\Rtwohm < 10$. \vspace{0.5em}

    \item We construct ACT and DES subsamples with similar mass and redshift distributions and find their mean SZ profiles to be consistent (Figure \ref{fig:ClusterSelection}). This implies that for clusters of a higher mass, $\langle \Mtwohm \rangle = 10^{14.85}\msol$, the SZ and optical selection effects do not amplify/suppress the deficit feature, and this adds to the robustness of the pressure deficit found in the ACTxACT and SPTxSPT measurements. \vspace{0.5em}

    \item For optically selected clusters of lower masses, $\Mtwohm < 10^{14.5} \msol$, the radial location of the log-derivative minima are consistent at $\Rad = 1.1$ while the depth becomes shallower with decreasing mass (Figure \ref{fig:GroupScales}).\vspace{0.5em}
    
    \item The SZ profiles measured around group-scale halos also differ significantly from the theoretical model in the one-halo regime, and are consistent with the model for the two-halo regime (Figure \ref{fig:GroupScales}). We discuss three potential causes for this: (i) mass estimation biases, (ii) deviations from the model of \citet{Battaglia2012PressureProfiles}, (iii) a non-zero correlation in the richness and SZ scatter at fixed mass. All three are more prominent for low-mass clusters, and we find none provide a clear explanation of the observations.\vspace{0.5em}
    
    \item We perform an oriented stacking of the clusters --- with the orientation determined by (i) the large-scale density field comprised of the surrounding structure, (ii) the brightest cluster galaxy, and (iii) a 2D Gaussian fit to the SZ image --- and split the profiles into three regions closest-to-furthest from the major axis. When using the LSS orientations, the two-halo term amplitude increases towards the major axis, while the log-derivative depth is steeper along the minor axis (Figure \ref{fig:OrientationChoice}). \vspace{0.5em}
    
    \item The location of the pressure deficit, $\Rshock$, is consistent with the splashback radius measured with galaxy number density profiles in  \citet{Shin2021SplashbackDESxACT}. The ratio is $\Rsp/\Rshock = 1.17 \pm 0.20$, and this consistency between shock and splashback radii further signifies the variety of dynamical processes happening at $R \approx \Rtwohm$. 
    
\end{itemize}

Our work, through the use of multiple independent datasets, shows the robustness of the pressure deficit feature in the outskirts of galaxy clusters. While we have discussed this feature as arising from the temperature difference between ions and electrons induced by shock heating, other physical processes could potentially cause this difference. The best way to identify the source of the feature is to obtain the electron number density and electron temperature profiles around these clusters. However, this is quite challenging given the deficit is in the outskirts of the cluster, and it is not possible for X-ray observations --- which are the primary way to measure these profiles --- to probe these regions for a large enough sample of clusters. Instead, it may be more possible to use high-fidelity X-ray observations of nearby individual clusters to look for such shocks in a small sample of low-redshift clusters.

Compared to \citetalias{Anbajagane2022Shocks}, we have focused less on the accretion shock feature in this work. While the ACT data shows some potential signs of a feature consistent with SPT data, the amplitude of the signal --- and thus the significance of the feature --- is low. This is not particularly surprising in that accretion shocks are highly irregular, in both their radial location around the clusters as well as in their geometry \citep{Zhang2020MergerAcceleratedShocks, Zhang2021SplashShock}. In fact, the simulation-based work of \citet{Baxter2021ShocksSZ} found the signal was clearly seen only when selecting relaxed clusters alone. To further pursue a detection of this feature, we can either redo our analyses with the release of a larger cluster catalog and/or lower-noise SZ maps, or perform selection cuts on the current catalogs (particulary related to cluster relaxation) that can maximize the signal-to-noise of this feature.

Moving forward there are still additional ongoing and future surveys/datasets that could be used for this work --- such as  SPT-3G \citep{Benson2014SPT3G}, Simons Observatory \citep{SimonsObs2019}, and CMB-S4 \citep{CMBS42019} --- that will all either have higher sensitivity and/or a larger sample of clusters across a broader range in mass and redshift. This would allow the study of the pressure deficit across redshift and mass. Finding clear trends may shed some light on the physical origin of the features. Combining existing datasets can also provide maps with a depth comparable to the upcoming CMB-S4 experiment, and the corresponding cluster catalogs --- such as the SPT Megadeep catalog \citepinprep{Kornoelje} --- will be particularly relevant for comparing the profile outskirts of low mass SZ-selected and optically selected clusters.

From the optical survey side, we have the cluster samples observed in the Kilo-degree survey \citep{Maturi2019KiDSClusters} and the Dark energy spectroscopic instrument legacy imaging survey \citep{Zou2021DESICluster} using richness selection techniques like in DES (but with different algorithms), and samples observed in Hyper Suprime-Cam using a weak-lensing mass selection \citep{Miyazaki2018ShearSelected, Chen2020ShearSelected}. The Hyper Suprime-Cam sample in particular accesses much higher redshifts than the DES dataset. Recently, the sample of X-ray selected clusters has also grown considerably, in part due to the eROSITA All-sky X-ray mission \citep{Liu2022erosita}. While X-ray samples have significantly lower redshift than the SZ and optical samples, they allow the pursuit of unique science cases --- X-ray clusters are bimodal in whether or not they have a cool core, and measuring the SZ profile outskirts around the two different types of clusters could shed light on the interplay between the physics of the outskirts and that of the cluster core. Opportunities also exist for studying the correlations between profiles, and these can have strong astrophysical signatures \citep[\eg][]{Farahi2022ProfileCorr}. Techniques have also been developed to extract such profile correlations in a data-driven manner, with minimal assumptions, such as Gaussian processes \citep{Farahi2021PoPE} and local linear regression \citep{Farahi2022KLLR}.

Thus, there are many synergistic opportunities for cross-correlating the different types of datasets --- both ongoing and upcoming --- and each combination will allow us to access different science cases regarding the physics of these cluster outskirts. The use of three independent, wide-field surveys in this work --- all analyzed under a common, coherent framework --- has given us the ability to easily cross-check and validate the signatures we see, and in general, be less sensitive to both known, and unknown, systematic effects. The ability to perform such tests and explorations will only grow, as we move into the age of even larger surveys with higher overlap and greater synergies. 

\section*{Acknowledgements}

DA is supported by NSF grant No. 2108168. CC is supported by the Henry Luce Foundation and DOE grant DE-SC0021949. MHi acknowledges support from the National Research Foundation of South Africa (grant no. 137975). KM acknowledges support from the National Research Foundation of South Africa. CS acknowledges support from the Agencia Nacional de Investigaci\'on y Desarrollo (ANID) through FONDECYT grant no.\ 11191125 and BASAL project FB210003. JCH acknowledges support from NSF grant AST-2108536, NASA grants 21-ATP21-0129 and 22-ADAP22-0145, the Sloan Foundation, and the Simons Foundation.

Funding for the DES Projects has been provided by the U.S. Department of Energy, the U.S. National Science Foundation, the Ministry of Science and Education of Spain, 
the Science and Technology Facilities Council of the United Kingdom, the Higher Education Funding Council for England, the National Center for Supercomputing 
Applications at the University of Illinois at Urbana-Champaign, the Kavli Institute of Cosmological Physics at the University of Chicago, 
the Center for Cosmology and Astro-Particle Physics at the Ohio State University,
the Mitchell Institute for Fundamental Physics and Astronomy at Texas A\&M University, Financiadora de Estudos e Projetos, 
Funda{\c c}{\~a}o Carlos Chagas Filho de Amparo {\`a} Pesquisa do Estado do Rio de Janeiro, Conselho Nacional de Desenvolvimento Cient{\'i}fico e Tecnol{\'o}gico and 
the Minist{\'e}rio da Ci{\^e}ncia, Tecnologia e Inova{\c c}{\~a}o, the Deutsche Forschungsgemeinschaft and the Collaborating Institutions in the Dark Energy Survey. 

The Collaborating Institutions are Argonne National Laboratory, the University of California at Santa Cruz, the University of Cambridge, Centro de Investigaciones Energ{\'e}ticas, 
Medioambientales y Tecnol{\'o}gicas-Madrid, the University of Chicago, University College London, the DES-Brazil Consortium, the University of Edinburgh, 
the Eidgen{\"o}ssische Technische Hochschule (ETH) Z{\"u}rich, 
Fermi National Accelerator Laboratory, the University of Illinois at Urbana-Champaign, the Institut de Ci{\`e}ncies de l'Espai (IEEC/CSIC), 
the Institut de F{\'i}sica d'Altes Energies, Lawrence Berkeley National Laboratory, the Ludwig-Maximilians Universit{\"a}t M{\"u}nchen and the associated Excellence Cluster Universe, 
the University of Michigan, NSF's NOIRLab, the University of Nottingham, The Ohio State University, the University of Pennsylvania, the University of Portsmouth, 
SLAC National Accelerator Laboratory, Stanford University, the University of Sussex, Texas A\&M University, and the OzDES Membership Consortium.

Based in part on observations at Cerro Tololo Inter-American Observatory at NSF's NOIRLab (NOIRLab Prop. ID 2012B-0001; PI: J. Frieman), which is managed by the Association of Universities for Research in Astronomy (AURA) under a cooperative agreement with the National Science Foundation.

The DES data management system is supported by the National Science Foundation under Grant Numbers AST-1138766 and AST-1536171.
The DES participants from Spanish institutions are partially supported by MICINN under grants ESP2017-89838, PGC2018-094773, PGC2018-102021, SEV-2016-0588, SEV-2016-0597, and MDM-2015-0509, some of which include ERDF funds from the European Union. IFAE is partially funded by the CERCA program of the Generalitat de Catalunya.
Research leading to these results has received funding from the European Research
Council under the European Union's Seventh Framework Program (FP7/2007-2013) including ERC grant agreements 240672, 291329, and 306478.
We  acknowledge support from the Brazilian Instituto Nacional de Ci\^encia
e Tecnologia (INCT) do e-Universo (CNPq grant 465376/2014-2).

This manuscript has been authored by Fermi Research Alliance, LLC under Contract No. DE-AC02-07CH11359 with the U.S. Department of Energy, Office of Science, Office of High Energy Physics.

The South Pole Telescope program is supported by the National Science Foundation (NSF) through the Grant No. OPP-1852617. Partial support is also provided by the Kavli Institute of Cosmological Physics at the University of Chicago.

Support for ACT was through the U.S.~National Science Foundation through awards AST-0408698, AST-0965625, and AST-1440226 for the ACT project, as well as awards PHY-0355328, PHY-0855887 and PHY-1214379. Funding was also provided by Princeton University, the University of Pennsylvania, and a Canada Foundation for Innovation (CFI) award to UBC. ACT operated in the Parque Astron\'omico Atacama in northern Chile under the auspices of the Agencia Nacional de Investigaci\'on y Desarrollo (ANID). The development of multichroic detectors and lenses was supported by NASA grants NNX13AE56G and NNX14AB58G. Detector research at NIST was supported by the NIST Innovations in Measurement Science program. Computing for ACT was performed using the Princeton Research Computing resources at Princeton University, the National Energy Research Scientific Computing Center (NERSC), and the Niagara supercomputer at the SciNet HPC Consortium. SciNet is funded by the CFI under the auspices of Compute Canada, the Government of Ontario, the Ontario Research Fund–Research Excellence, and the University of Toronto. We thank the Republic of Chile for hosting ACT in the northern Atacama, and the local indigenous Licanantay communities whom we follow in observing and learning from the night sky.

All analysis in this work was enabled greatly by the following software: \textsc{Pandas} \citep{Mckinney2011pandas}, \textsc{NumPy} \citep{vanderWalt2011Numpy}, \textsc{SciPy} \citep{Virtanen2020Scipy}, and \textsc{Matplotlib} \citep{Hunter2007Matplotlib}. We have also used
the Astrophysics Data Service (\href{https://ui.adsabs.harvard.edu/}{ADS}) and \href{https://arxiv.org/}{\texttt{arXiv}} preprint repository extensively during this project and the writing of the paper.

\section*{Data Availability}

All SPT data used in our analyses are publicly available at the repositories linked to in this paper. The cluster catalog from ACT DR5 is publicly available, alongside the maps from which these catalogs are constructed. The SZ map of ACT DR6, and the raw DR6 maps used to construct it, will be made public shortly. The DR4 maps used to construct this SZ map are already available. The DES Year 3 shape catalog is also available, while the cluster catalog is not yet public. The links to the online portals hosting the publicly available catalogs can be found under the relevant data subsection in Section \ref{sec:Data}.

The code used to generate the theoretical tSZ profile of a halo, including both one-halo and two-halo contributions, is made available at \url{https://github.com/DhayaaAnbajagane/tSZ_Profiles}.

\bibliographystyle{mnras}
\bibliography{References}

\appendix

\section{Impact of miscentering on profiles}\label{appx:Miscentering}

\begin{figure*}
    \centering
    \includegraphics[width = 2\columnwidth]{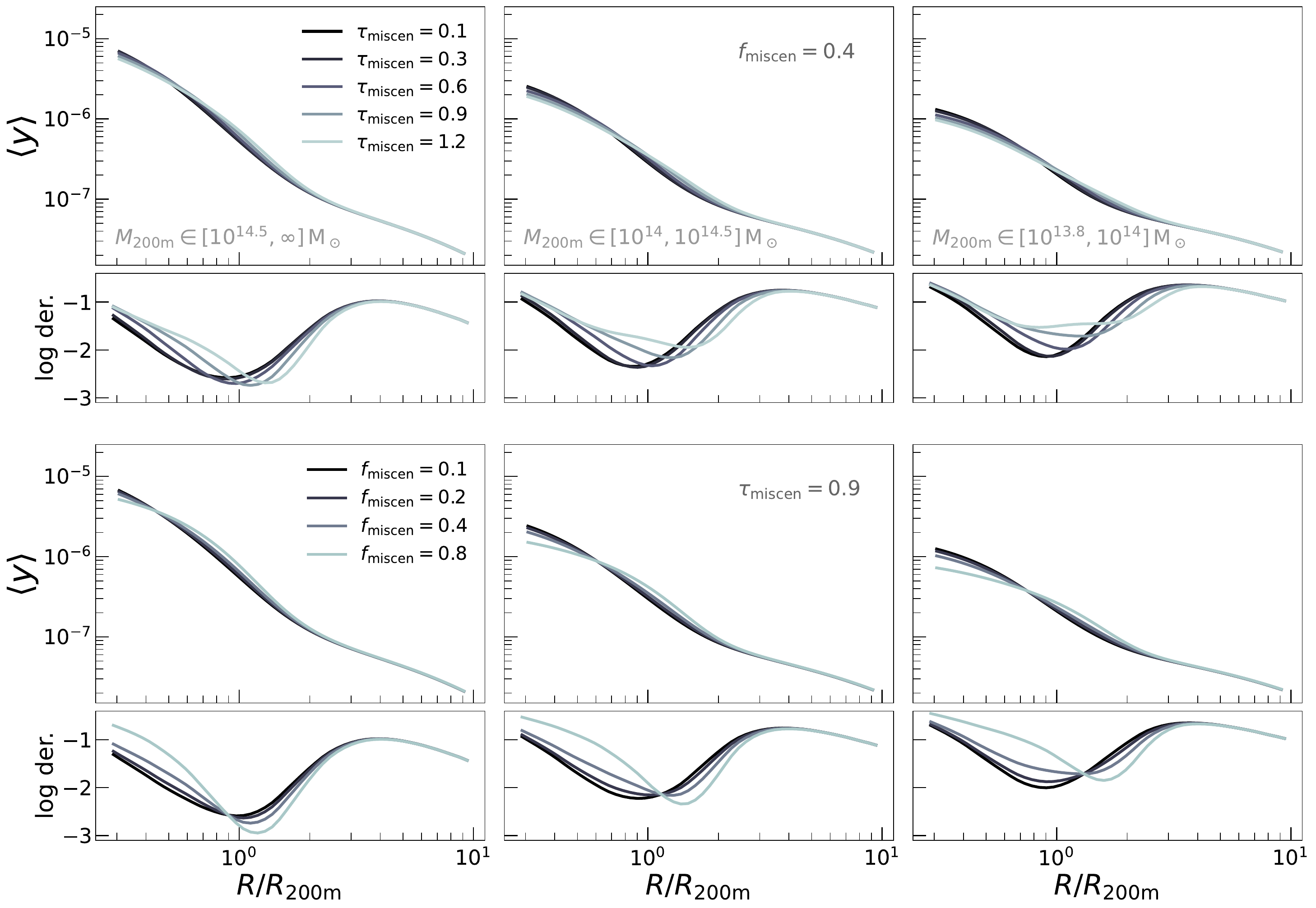}
    \caption{The DESxSPT theory predictions --- both mean profile (top) and log-derivative (bottom) --- when varying the parameters of the miscentering model. We vary $\tau_{\rm miscen}$, which is the miscentering length scale (top two panels), and miscentering fraction $f_{\rm miscen}$ (bottom two panels). The different columns show the model for different mass ranges. Miscentering transfers power from small-scales ($R < 0.3\Rtwohm$) to large-scales ($R \sim \Rtwohm$). This is distinct from the impacts of contamination, which will suppress the one-halo term over all scales, but similar to feedback which will push gas from small-scales out to large-scales.}
    \label{fig:Miscentering}
\end{figure*}

As we discussed previously, our theoretical model for the SZ profiles of \textit{optically selected} clusters depends on the miscentering model parameters assumed for the cluster sample. In Figure \ref{fig:Miscentering} we show the model for the DES clusters' SZ profile as we vary the amplitude of miscentering, $\tau_{\rm miscen}$, and the fraction of miscentered clusters, $f_{\rm miscen}$, defined in Section \ref{sec:MiscenteringModel}. We do not have an external, calibrated constraint for the miscentering effect in this specific DES cluster sample. Thus, as an alternative, we vary the parameter values until the theory visually matches the data for the one-halo term as shown in Figure \ref{fig:Profiles}. In practice, we do this by making predictions in a 5x5 grid of parameter values, and find $\tau_{\rm miscen} = 0.9$ and $f_{\rm miscen} = 0.4$ to be the best combination. As was noted before, both values are near the $3-4\sigma$ upper limit of constraints on the DES Y1 cluster sample, depending on the parameter \citep[][see their Chandra--DES constraints in Table 1]{Zhang2019MiscenteringDESY1}, while the value of $\tau_{\rm miscen}$ is within $1\sigma$ of the estimate from \citet[][see their Table 6]{Bleem2020SPT-ECS}, which is based on a SPT-DES matched cluster sample. It is also generally consistent with the work of \citet{Sehgal2013ACTMiscenter}, who find the offsets in individual clusters seen in ACT have upper limits of $1.5\mpc$, which corresponds to $\tau_{\rm miscen} \approx 1.5$.

Given these potential limitations of the implemented miscentering effects in our work, we focus our analysis of optically selected cluster on results that do not require accurate theoretical estimates of pressure profiles for these clusters. We specifically avoid quoting a detection significance of shock features in these clusters given the uncertainty in the miscentering model parameters of the theoretical model. Our results in Section \ref{sec:GroupScales}, which \textit{does} compare theory and data for low mass DES clusters and finds large deviations, are insensitive to miscentering as the deviations are significantly larger than those from miscentering effects alone.

Figure \ref{fig:Miscentering} shows that for the high mass sample (left panels), the variation in $\tau_{\rm miscen}$ changes the location of the log-derivative minimum from $r_{\rm min} = 0.7\Rtwohm \rightarrow 1.2 \Rtwohm$ as we vary $\tau_{\rm miscen} = 0.1 \rightarrow 1.2$. The minimum value of the log-derivative goes from $-2.5  \rightarrow -3.0$ as we vary $f_{\rm miscen} = 0.1 \rightarrow 0.8$. In particular, the result for $f_{\rm miscen} = 0.8$ and $\tau_{\rm miscen} = 0.9$ appears to replicate a shock-esque feature at $R \approx \Rtwohm$. However, this is not an indication that shock features can be explained by miscentering. For SZ-selected clusters, the miscentering is much smaller than than the values being considered in Figure \ref{fig:Miscentering}. While the predicted profile for large miscentering values forms a deficit-like feature, the agreement in the one-halo term is significantly degraded as a result. Thus, this is not evidence that miscentering is the cause of the deficit feature, and is instead evidence of the miscentering model's ability to capture steep drops in the pressure profile.

\section{Dependence on cluster SNR} \label{appx:SNRDependence}

\begin{figure*}
    \centering
    \includegraphics[width = 1.6\columnwidth]{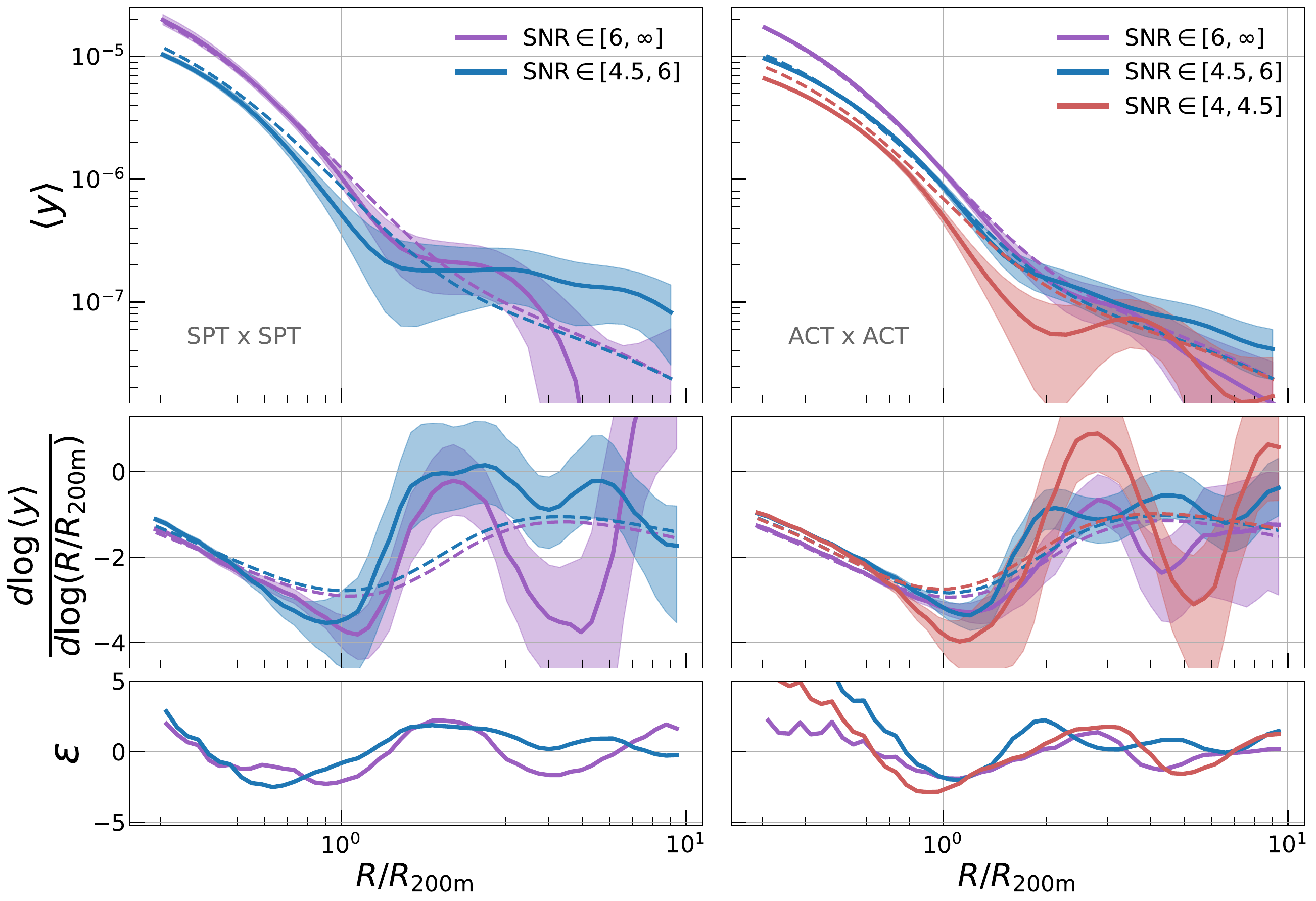}
    \caption{The average SZ profiles of SZ-selected clusters measured on their respective SZ maps, with the samples split by their signal-to-noise ratio. All subsamples show pressure deficits, confirming that there is no SNR dependence on the deficit feature. The location and depth of the log-derivative minima are listed in Table \ref{tab:Results}.}
    \label{fig:SNRDependence}
\end{figure*}

The analysis in Section \ref{sec:SurveyComparisons} implies the pressure deficit is not formed due to SZ selection effects. A characteristic of a feature driven by noise-effects is its amplitude grows near the limit of the selection threshold. For SZ-selected clusters, this is the signal-to-noise threshold, which is $\texttt{SNR} > 4.5$ for ACT and $\texttt{SNR} > 4$ for SPT. In Figure \ref{fig:SNRDependence} we show the average SZ profiles of cluster subsamples split by their SNR. The pressure deficit feature exists in all SNR bins, and the values for the location and log-derivative of the deficit (listed in Table \ref{tab:Results}) are consistent within $<1\sigma$. This adds further to the evidence that the feature is not formed from SZ noise-based selection effects.

\section{Analytic fit for pressure deficit}\label{appx:SamFits}

\begin{figure}
    \centering
    \includegraphics[width = \columnwidth]{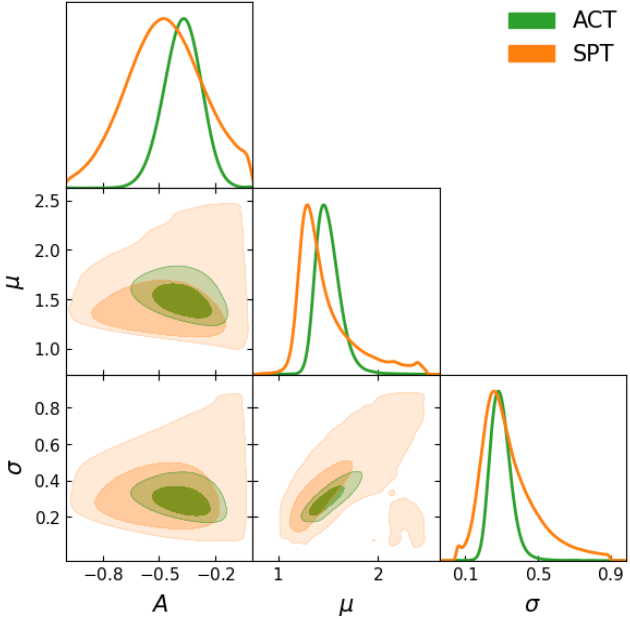}
    \caption{The constraints on the parameters of Equation \eqref{eq:gauss} obtained from the SPT or ACT data. The values are listed in Table \ref{tab:Fit_Constraints}. The constraints from SPT and ACT are consistent with each other within $<0.5 \sigma$, which independently validates our statement that the  pressure deficit features shown in Figure \ref{fig:Profiles} for these samples is consistent.}
    \label{fig:TrianglePlot}
\end{figure}

\begin{figure}
    \centering
    \includegraphics[width = \columnwidth]{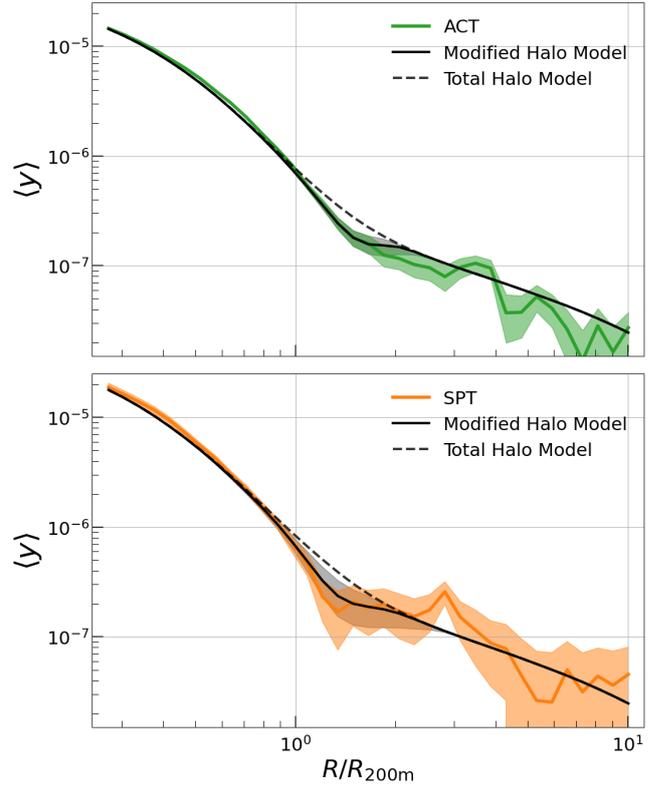}
    \caption{The unsmoothed average SZ profile of the ACT sample (top) and SPT sample (bottom). Overlaid is the original theory of the total halo model in black dashed line, and the modified halo model, described in Equation \eqref{eq:gauss}, in the solid black line. The modified model provides a better fit to the data, particularly to the pressure deficit term. Note that the original halo model is a prediction and not a fit, whereas the modified halo theory fits the additional Gaussian component while keeping the original halo theory component fixed.}
    \label{fig:ProfilePlot}
\end{figure}

\begin{table}
    \centering
    \begin{tabular}{c|c|c|c|c|c}
        Dataset & $A$ & $\mu$ & $\sigma$ & $\chi^2_{\rm mod}$ & $\chi^2_{\rm orig}$\\
        \hline 
        \hline
        SPT x SPT & $-0.47^{+0.21}_{-0.21}$ & $1.39^{+0.36}_{-0.15}$ & $0.31^{+0.18}_{-0.10}$ & $23.41$ & $32.86$ \\ \\
        ACT x ACT & $-0.37^{+0.10}_{-0.10}$ & $1.49^{+0.13}_{-0.10}$ & $0.29^{+0.06}_{-0.05}$ & $65.97$ & $86.43$\\
        \hline
    \end{tabular}
    \caption{The best-fit parameters of the modified halo model ($A, \mu, \sigma$) for both SPT and ACT data. The $\chi^2_{\rm orig}$ and $\chi^2_{\rm mod}$ columns are the chi-squared for the original and modified halo models. The modified model is significantly better in both cases.}
    \label{tab:Fit_Constraints}
\end{table}

In Figure \ref{fig:Profiles}, the halo model is a good match to the measured profiles in both the one-halo and two-halo regimes, but the pressure deficit feature in the transition region is only seen in the measurements. We implement a simple modification to the existing halo model theory to match this effect. Our modification multiplies the original theory by a Gaussian,
\begin{equation} \label{eq:gauss}
    \xi_{h,p}^{\rm mod}(r,A,\mu,\sigma) = \xi_{h,p} \times \left(1 + A \frac{\mathscr{N}(r, \mu, \sigma)}{\mathscr{N}(\mu, \mu, \sigma)} \right),
\end{equation}
where $A$ is the amplitude of the Gaussian, $\mu$ is the mean/location in units of $R/\Rtwohm$, $\sigma$ is the width of the Gaussian, $r$ is the comoving distance bins mentioned previously in Section \ref{sec:Model_detection_significance}, and $\xi_{h,p}$ is the halo-pressure correlation computed in that same section, accounting for beam-smoothing effects. Equation \eqref{eq:gauss} shows that we additionally normalize the Gaussian feature, $\mathscr{N}(r, \mu, \sigma)$, by another Gaussian evaluated at the mean, $\mathscr{N}(\mu, \mu, \sigma)$, and this ensures that only the parameter $A$ controls the amplitude of the Gaussian.\footnote{Without this renormalization, the parameter $\sigma$ would also control the amplitude in addition to $A$, given the Gaussian goes as $\mathcal{N} \propto 1/\sqrt{2\pi\sigma^2}$. This then causes degeneracies in the parameter space that lead to problems in the bayesian inference.}

We then fit the model in Equation \eqref{eq:gauss} to the measured profiles and obtain constraints on the three parameters, $A$, $\mu$ and $\sigma$. The fit is done using the Markov Chain Monte Carlo (MCMC) technique as implemented in the \textsc{Emcee} package \citep{emcee}. We use a set of mostly uninformative priors for all parameters,
\begin{align}\label{eqn:priors}
    -10 < A &< 0\nonumber\\
    0.8 < \mu &< 2.5\\
    0.05 < \sigma &<0.9\nonumber.
\end{align}
The bounds of $\mu$, and $\sigma$ are chosen to prevent the fitting of either random noise fluctuations in the profiles or differences between data and theory near the cluster core. We enforce $A < 0$ as we are fitting a deficit feature and so the Gaussian must suppress (and not amplify) the pressure in the existing theory prediction. The fit is performed using only $R > 0.6\Rtwohm$, and this is done primarily to prevent the MCMC from focusing on any deviations between data and theory in the cluster core. Limiting our analysis to this radial range also helps obtain a numerically stable covariance matrix for use in the MCMC. The fitting is performed by minimizing the $\chi^2$ for the log-likelihood:
\begin{equation}
    \chi^2 = \left(y^{\rm{obs}}-y^{\rm{th}}\right) C^{-1} \left(y^{\rm{obs}}-y^{\rm{th}}\right),
\end{equation}
where $C$ is the covariance matrix of the profile, $y^{\rm{obs}}$ is the measured profile, and $y^{\rm{th}}$ is the modified halo model. The MCMC is run on the raw, unsmoothed profiles using 300 walkers and 40,000 steps per walker. 

We show the results for the SPT and ACT cluster catalogs, each measured on their corresponding SZ maps. The parameters corresponding to this fit are shown in Figure \ref{fig:TrianglePlot}, while the fits and the data are compared in Figure \ref{fig:ProfilePlot}. The latter shows that the modified halo model is a good fit to the data, and better than the original total halo model, in the one-to-two halo transition regime with the pressure deficit feature. The constraints from the ACT and SPT samples shown in Figure \ref{fig:TrianglePlot} are consistent with each other. Table \ref{tab:Fit_Constraints} lists the amplitude, size, and location of the pressure deficit, and we see the values are consistent within $<0.5\sigma$ across the two samples. We also list the $\chi^2$ of the fit from the original halo model and the modified halo model. For both datasets, the $\chi^2$ is noticeably improved, and we can also see this visually in the fits of Figure \ref{fig:ProfilePlot}. These fits can be used as a simple technique to include SPT/ACT-like pressure deficits in an existing halo model.

In our main analysis, we have not performed any fits, which has primarily been due to the lack of a model for the pressure deficit. In this section, we now show we do have such a model. However, it is not used in our main analysis as we have not yet studied its robustness and validated it against any potential biases. For example, we have already found that it is fairly easy for this model to fit features other than the pressure deficit  --- like fluctuations on small scales --- and the priors must be slightly hand-tuned to make the model focus on the deficit. In our case, the bounds on $\sigma$ and $\mu$ were tuned so as to avoid such issues when fitting the two mean SZ profiles we present here. It is unlikely these priors can be used generically for all measured mean profiles without running into fit failures or prior boundary effects. Thus, while the fits we describe here are a useful phenomenological model, they have not been validated at the same rigor as our current pipeline (which was tested extensively in \citetalias{Anbajagane2022Shocks}) and so we have continued with the original pipeline for the main analysis in this work.

In the future, one could also use this technique --- namely, the posteriors of the model parameters --- as an alternative estimator of the detection significance for the pressure deficit, where $A = 0$ would denote no detection of the deficit. While the results of Table \ref{tab:Fit_Constraints} already provide the relevant numbers for this work, we do not quote this detection significance as we have yet to validate our profile-fitting technique adequately. Thus, the main purpose and result of this section remain the fits that enable a simple, data-driven replication of the deficit feature in a halo model prediction.

\section{Correlation matrix}\label{sec:CorrMat}

\begin{figure}
    \centering
    \includegraphics[width = \columnwidth]{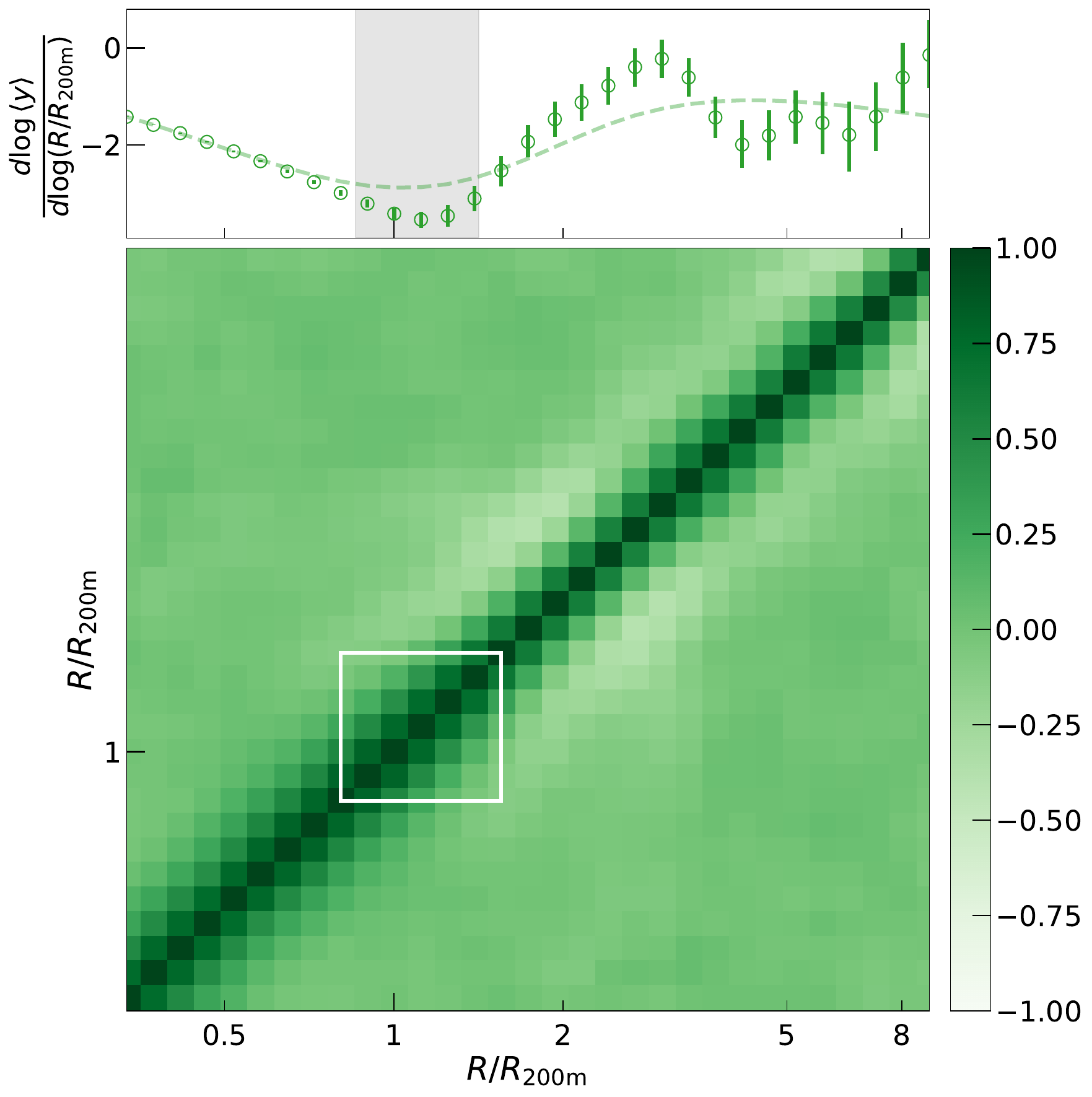}
    \caption{The top panel shows the log-derivative for the ACTxACT measurement (presented in Figure \ref{fig:Profiles}), now presented as discrete points corresponding to the measurements within each bin. The dashed line is the theoretical prediction, and the gray band shows the range of scales used to quantify the detection of a pressure deficit. The bottom panel shows the correlation matrix of the log-derivative. The matrices for the other measurements have the same structure. The white box demarcates the bins used in computing the $\chi^2$ of the pressure deficit, as listed in Table \ref{tab:Results}.}
    \label{fig:CorrMat}
\end{figure}

Figure \ref{fig:CorrMat} presents the correlation matrix of the ACTxACT log-derivative measurements. It is a typical diagonal matrix, with some correlations in adjacent bins due to the smoothing procedure (see Section \ref{sec:Measurement}). The white box highlights the bins used to estimate the $\chi^2$ shown in Table \ref{tab:Results}. The top panel shows the log-derivative measurement, now in discrete points corresponding to the binning, corresponding to the ACTxACT measurement of Figure \ref{fig:Profiles}.

\section{Impact of Cosmic Infrared Background contamination}\label{sec:CIB}

\begin{figure}
    \centering
    \includegraphics[width = \columnwidth]{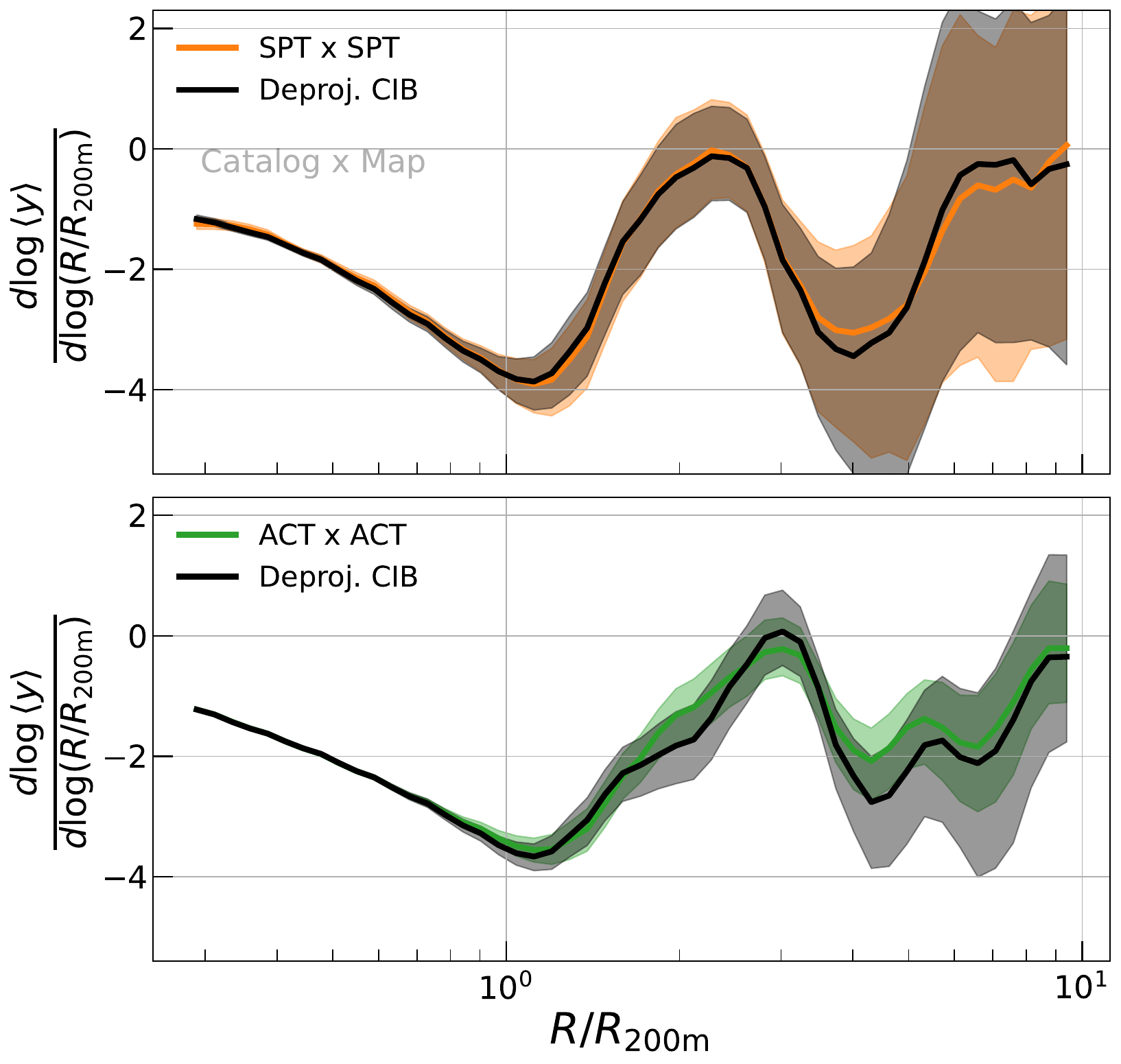}
    \caption{The log-derivatives of the SPTxSPT and ACTxACT datasets, measured using the fiducial maps (colored lines) and maps with the CIB deprojected (black lines). The impact of the CIB on this measurement is negligible as the two versions are statistically consistent across the range of scales.}
    \label{fig:CIBTest}
\end{figure}

We explicitly test the impact of the cosmic infrared background (CIB) on our SZ profile measurements by comparing the fiducial maps with those where the CIB signal is deprojected/minimized in the final maps. See \citet{Bleem2021SPTymap, Coulton2023ACTyMaps} for details on the deprojection procedure of SPT and ACT, respectively. In Figure \ref{fig:CIBTest} we compare measurements of the log-derivative made on these two sets of maps, for both SPT and ACT. The measurements are statistically consistent across the maps with and without CIB deprojection. The fiducial SPT map already removes a significant fraction of the CIB, as discussed in \citet[][see their Section 3.5]{Bleem2021SPTymap}. The ACT data contain multiple CIB-deprojected maps, and we use the fiducial one \citep[][see their Section C.3 and Equation 18]{Coulton2023ACTyMaps}.

\section{Affiliations}

$^{1}$ Department of Astronomy and Astrophysics, University of Chicago, Chicago, IL 60637, USA\\
$^{2}$ Kavli Institute for Cosmological Physics, University of Chicago, Chicago, IL 60637, USA\\
$^{3}$ Institute for Astronomy, University of Hawai'i, 2680 Woodlawn Drive, Honolulu, HI 96822, USA\\
$^{4}$ Department of Physics and Astronomy, University of Pennsylvania, Philadelphia, PA 19104, USA\\
$^{5}$ David A. Dunlap Department of Astronomy \& Astrophysics, University of Toronto, 50 St. George St., Toronto, ON M5S 3H4, Canada\\
$^{6}$ Canadian Institute for Theoretical Astrophysics, University of Toronto, 60 St. George St., Toronto, ON M5S 3H4, Canada\\
$^{7}$ Dunlap Institute of Astronomy \& Astrophysics, 50 St. George St., Toronto, ON M5S 3H4, Canada\\
$^{8}$ Laborat\'orio Interinstitucional de e-Astronomia - LIneA, Rua Gal. Jos\'e Cristino 77, Rio de Janeiro, RJ - 20921-400, Brazil\\
$^{9}$ Fermi National Accelerator Laboratory, P. O. Box 500, Batavia, IL 60510, USA\\
$^{10}$ Department of Physics, University of Michigan, Ann Arbor, MI 48109, USA\\
$^{11}$ Institute of Astronomy, University of Cambridge, Madingley Road, Cambridge CB3 0HA, UK\\
$^{12}$ Kavli Institute for Cosmology, University of Cambridge, Madingley Road, Cambridge CB3 0HA, UK\\
$^{13}$ Department of Physics and Astronomy, University of Southern California, Los Angeles, CA 90089, USA\\
$^{14}$ Institute of Cosmology and Gravitation, University of Portsmouth, Portsmouth, PO1 3FX, UK\\
$^{15}$ Department of Astronomy, Cornell University, Ithaca, NY 14853, USA\\
$^{16}$ Physics Department, 2320 Chamberlin Hall, University of Wisconsin-Madison, 1150 University Avenue Madison, WI  53706-1390\\
$^{17}$ Argonne National Laboratory, 9700 South Cass Avenue, Lemont, IL 60439, USA\\
$^{18}$ Fermi National Accelerator Laboratory, MS209, P.O. Box 500, Batavia, IL 60510\\
$^{19}$ High Energy Physics Division, Argonne National Laboratory, Argonne, IL, USA 60439\\
$^{20}$ Faculty of Physics, Ludwig-Maximilians-Universitat, Scheinerstr. 1, 81679 Munich, Germany\\
$^{21}$ Canadian Institute for Theoretical Astrophysics, University of Toronto, Toronto, ON, Canada M5S 3H8\\
$^{22}$ Department of Physics \& Astronomy, University College London, Gower Street, London, WC1E 6BT, UK\\
$^{23}$ Instituto de Astrofisica de Canarias, E-38205 La Laguna, Tenerife, Spain\\
$^{24}$ Universidad de La Laguna, Dpto. Astrofísica, E-38206 La Laguna, Tenerife, Spain\\
$^{25}$ Department of Astronomy, University of Illinois at Urbana-Champaign, 1002 W. Green Street, Urbana, IL 61801, USA\\
$^{26}$ Center for Astrophysical Surveys, National Center for Supercomputing Applications, 1205 West Clark St., Urbana, IL 61801, USA\\
$^{27}$ Department of Physics, Duke University Durham, NC 27708, USA\\
$^{28}$ NASA Goddard Space Flight Center, 8800 Greenbelt Rd, Greenbelt, MD 20771, USA\\
$^{29}$ Astronomy Unit, Department of Physics, University of Trieste, via Tiepolo 11, I-34131 Trieste, Italy\\
$^{30}$ INAF-Osservatorio Astronomico di Trieste, via G. B. Tiepolo 11, I-34143 Trieste, Italy\\
$^{31}$ Institute for Fundamental Physics of the Universe, Via Beirut 2, 34014 Trieste, Italy\\
$^{32}$ Institute of Space Sciences (ICE, CSIC),  Campus UAB, Carrer de Can Magrans, s/n,  08193 Barcelona, Spain\\
$^{33}$ Institut d'Estudis Espacials de Catalunya (IEEC), 08034 Barcelona, Spain\\
$^{34}$ Hamburger Sternwarte, Universit\"{a}t Hamburg, Gojenbergsweg 112, 21029 Hamburg, Germany\\
$^{35}$ School of Mathematics and Physics, University of Queensland,  Brisbane, QLD 4072, Australia\\
$^{36}$ Centro de Investigaciones Energ\'eticas, Medioambientales y Tecnol\'ogicas (CIEMAT), Madrid, Spain\\
$^{37}$ Department of Physics, IIT Hyderabad, Kandi, Telangana 502285, India\\
$^{38}$ Universit\'e Grenoble Alpes, CNRS, LPSC-IN2P3, 38000 Grenoble, France\\
$^{39}$ Department of Physics and Astronomy, University of Waterloo, 200 University Ave W, Waterloo, ON N2L 3G1, Canada\\
$^{40}$ Institute of Theoretical Astrophysics, University of Oslo. P.O. Box 1029 Blindern, NO-0315 Oslo, Norway\\
$^{41}$ SLAC National Accelerator Laboratory, Menlo Park, CA 94025, USA\\
$^{42}$ Instituto de Fisica Teorica UAM/CSIC, Universidad Autonoma de Madrid, 28049 Madrid, Spain\\
$^{43}$ Institut de F\'{\i}sica d'Altes Energies (IFAE), The Barcelona Institute of Science and Technology, Campus UAB, 08193 Bellaterra (Barcelona) Spain\\
$^{44}$ Universit\"at Innsbruck, Institute f\"ur Astro- und Teilchenphysik, Technikerstrasse 25/8, 6020 Innsbruck, Austria\\
$^{45}$ University Observatory, Faculty of Physics, Ludwig-Maximilians-Universit\"at, Scheinerstr. 1, 81679 Munich, Germany\\
$^{46}$ School of Physics and Astronomy, Cardiff University, CF24 3AA, UK\\
$^{47}$ Department of Physics, Columbia University, 538 West 120th Street, New York, NY, USA 10027\\
$^{48}$ Wits Centre for Astrophysics, School of Physics, University of the Witwatersrand, Private Bag 3, 2050, Johannesburg, South Africa\\
$^{49}$ Astrophysics Research Centre, School of Mathematics, Statistics, and Computer Science, University of KwaZulu-Natal, Westville Campus, Durban 4041, South Africa\\
$^{50}$ Santa Cruz Institute for Particle Physics, Santa Cruz, CA 95064, USA\\
$^{51}$ Department of Physics, The Ohio State University, Columbus, OH 43210, USA\\
$^{52}$ Center for Cosmology and Astro-Particle Physics, The Ohio State University, Columbus, OH 43210, USA\\
$^{53}$ Center for Astrophysics $\vert$ Harvard \& Smithsonian, 60 Garden Street, Cambridge, MA 02138, USA\\
$^{54}$ Australian Astronomical Optics, Macquarie University, North Ryde, NSW 2113, Australia\\
$^{55}$ Lowell Observatory, 1400 Mars Hill Rd, Flagstaff, AZ 86001, USA\\
$^{56}$ Department of Applied Mathematics and Theoretical Physics, University of Cambridge, Cambridge CB3 0WA, UK\\
$^{57}$ George P. and Cynthia Woods Mitchell Institute for Fundamental Physics and Astronomy, and Department of Physics and Astronomy, Texas A\&M University, College Station, TX 77843,  USA\\
$^{58}$ Kavli Institute for Particle Astrophysics \& Cosmology, P. O. Box 2450, Stanford University, Stanford, CA 94305, USA\\
$^{59}$ Department of Physics, University of Chicago, Chicago, IL 60637, USA\\
$^{60}$ Enrico Fermi Institute, University of Chicago, Chicago, IL 60637, USA\\
$^{61}$ Instituci\'o Catalana de Recerca i Estudis Avan\c{c}ats, E-08010 Barcelona, Spain\\
$^{62}$ Astrophysics Research Centre, University of KwaZulu-Natal, Westville Campus, Durban 4041, South Africa\\
$^{63}$ European Southern Observatory, Karl-Schwarzschild-Str. 2, Garching 85748, Germany\\
$^{64}$ Department of Physics, Stanford University, 382 Via Pueblo Mall, Stanford, CA 94305, USA\\
$^{65}$ Institute of Theoretical Astrophysics, University of Oslo, Norway\\
$^{66}$ Instituto de F\'isica Gleb Wataghin, Universidade Estadual de Campinas, 13083-859, Campinas, SP, Brazil\\
$^{67}$ Observat\'orio Nacional, Rua Gal. Jos\'e Cristino 77, Rio de Janeiro, RJ - 20921-400, Brazil\\
$^{68}$ Joseph Henry Laboratories of Physics, Jadwin Hall, Princeton University, Princeton, NJ, USA 08544\\
$^{69}$ Department of Physics, Carnegie Mellon University, Pittsburgh, Pennsylvania 15312, USA\\
$^{70}$ Department of Physics and Astronomy, Haverford College, Haverford, PA, USA 19041\\
$^{71}$ Ruhr University Bochum, Faculty of Physics and Astronomy, Astronomical Institute (AIRUB), German Centre for Cosmological Lensing, 44780 Bochum, Germany\\
$^{72}$ Institute for Astronomy, University of Edinburgh, Edinburgh EH9 3HJ, UK\\
$^{73}$ School of Physics, University of Melbourne, Parkville, VIC 3010, Australia\\
$^{74}$ Laboratoire de physique des 2 infinis Ir\`ene Joliot-Curie, CNRS Universit\'e Paris-Saclay, B\^at. 100, Facult\'e des sciences, F-91405 Orsay Cedex, France\\
$^{75}$ Jodrell Bank Center for Astrophysics, School of Physics and Astronomy, University of Manchester, Oxford Road, Manchester, M13 9PL, UK\\
$^{76}$ Department of Physics and Astronomy, Pevensey Building, University of Sussex, Brighton, BN1 9QH, UK\\
$^{77}$ Kavli Institute for Particle Astrophysics and Cosmology and Department of Physics, Stanford University, Stanford, CA 94305, USA\\
$^{78}$ Brookhaven National Laboratory, Bldg 510, Upton, NY 11973, USA\\
$^{79}$ Department of Physics and Astronomy, Stony Brook University, Stony Brook, NY 11794, USA\\
$^{80}$ Instituto de F\'isica, Pontificia Universidad Cat\'olica de Valpara\'iso, Casilla 4059, Valpara\'iso, Chile\\
$^{81}$ School of Physics and Astronomy, University of Southampton,  Southampton, SO17 1BJ, UK\\
$^{82}$ Computer Science and Mathematics Division, Oak Ridge National Laboratory, Oak Ridge, TN 37831\\
$^{83}$ National Center for Supercomputing Applications, 1205 West Clark St., Urbana, IL 61801, USA\\
$^{84}$ Institut de Recherche en Astrophysique et Plan\'etologie (IRAP), Universit\'e de Toulouse, CNRS, UPS, CNES, 14 Av. Edouard Belin, 31400 Toulouse, France\\
$^{85}$ Department of Physics, Cornell University, Ithaca, NY, 14853, USA\\
$^{86}$ Lawrence Berkeley National Laboratory, 1 Cyclotron Road, Berkeley, CA 94720, USA\\
$^{87}$ Max Planck Institute for Extraterrestrial Physics, Giessenbachstrasse, 85748 Garching, Germany\\
$^{88}$ Universit\"ats-Sternwarte, Fakult\"at f\"ur Physik, Ludwig-Maximilians Universit\"at M\"unchen, Scheinerstr. 1, 81679 M\"unchen, Germany\\
$^{89}$ NASA/Goddard Space Flight Center, Greenbelt, MD, USA 20771\\

\bsp	
\label{lastpage}
\end{document}